\documentclass[12pt]{article}

\usepackage[letterpaper, margin=1in]{geometry}
\usepackage{setspace}
\usepackage{amsmath,amssymb,amsfonts}
\usepackage{enumitem}
\usepackage{booktabs}
\usepackage{longtable}
\usepackage{array}
\usepackage{amsmath}
\usepackage{longtable}
\usepackage{booktabs}
\usepackage{array}
\usepackage{ragged2e}
\usepackage{xurl}
\usepackage{placeins}
\usepackage{amssymb}

\newcommand{\feature}[1]{\nolinkurl{#1}}

\usepackage{graphicx}
\usepackage{booktabs}
\usepackage{array}
\usepackage{multirow}
\usepackage{float}
\usepackage[caption=false,font=footnotesize]{subfig}
\usepackage{tikz}
\usetikzlibrary{positioning}
\usepackage{graphicx}
\usepackage{tabularx}

\usepackage{titlesec}

\titleformat{\paragraph}
  {\normalfont\normalsize\bfseries}
  {\theparagraph}
  {1em}
  {}

\titlespacing*{\paragraph}
  {0pt}
  {3.25ex plus 1ex minus 0.2ex}
  {1.5ex plus 0.2ex}

\usepackage{listings}
\usepackage{xcolor}

\usepackage[round,authoryear]{natbib}
\usepackage{url}
\usepackage[colorlinks=true, linkcolor=black, citecolor=black, urlcolor=black]{hyperref}

\usepackage{authblk}

\newcommand{\Unknown}{\textsc{Unknown}}

\title{
\vspace{-2.0cm}
\Large
Directed-Hypergraph Analytics and Fragility Testing \\for Anti-Money Laundering
}
\author[1]{Run Li\thanks{Corresponding author. E-mail: tonirli@udel.edu. Funding: This research received no specific grant from any funding agency in the public, commercial, or not-for-profit sectors.}}
\author[2]{Paul Laux}

\affil[1]{Institute for Financial Services Analytics, University of Delaware, Newark, DE 19716, USA. E-mail: tonirli@udel.edu}
\affil[2]{Department of Finance and Institute for Financial Services Analytics, University of Delaware, Newark, DE 19716, USA. E-mail: laux@udel.edu}

\date{}

\begin{document}

\maketitle

\vspace{-1.2em}

\begin{abstract}
\noindent Illicit activity in financial networks is difficult to detect. We exploit the idea that the chance of a transaction being illicit depends partly on the concurrent positions and recent actions of counterparties, as well as the higher-order organization of fund flows. We combine hypergraph features and machine learning to demonstrate practical effectiveness. Specifically, we use the Elliptic++ Bitcoin dataset to show that graph- and directed-hypergraph-based features improve illicit-address classification. The success of the predictive model then allows us to examine potential money-laundering chains are structurally fragile. Several illicit-to-licit transaction motifs with unknown intermediary addresses specified, and the fragility of these motifs is evaluated through progressive removal of the highest-risk intermediaries. Surprisingly, these motifs are fragile, suggesting that money laundering could be disrupted via the identification and exclusion of owners of a limited number of crypto wallets.
\end{abstract}

\vspace{0.5em}

\section{Introduction}
\label{sec:introduction}

Money laundering is the process through which proceeds generated from illegal activity are transferred, transformed, or concealed so that their criminal origin becomes more difficult to identify and the funds can eventually appear legitimate. It is commonly associated with fraud, corruption, drug trafficking, cybercrime, organized crime, and other predicate offenses. A central feature of money laundering is that illicit funds are rarely moved directly from a criminal source to a final legitimate destination. Instead, they may pass through multiple accounts, intermediaries, transactions, or financial services in a process called ``layering'' to obscure their origin. The United Nations Office on Drugs and Crime estimates that the amount of money laundered globally each year may be equivalent to approximately 2--5\% of global GDP \citep{unodc2011moneylaundering}. 

The economic scale of the problem has made anti-money-laundering (AML) enforcement an important responsibility of governments, regulators, law-enforcement agencies, and financial institutions. International standards established by the Financial Action Task Force require financial institutions to implement risk-based procedures for identifying and monitoring potentially suspicious activity \citep{fatf2012recommendations}. Substantial resources are devoted to AML compliance, transaction monitoring, customer due diligence, suspicious-activity reporting, and financial-crime investigations. 

However, conventional AML systems face an important statistical difficulty: financial institutions process extremely large numbers of legitimate transactions while truly illicit activities are rare. In addition, suspicious behavior may be distributed across several accounts and transactions rather than concentrated in a single observation. Improving the identification of illicit financial activity therefore has economic value not only because it can increase detection accuracy, but also because it can help regulators and financial institutions allocate costly investigative resources more efficiently.

This problem is inherently network-based. An individual transaction that appears ordinary in isolation may become informative when its position within a broader fund-flow network is considered. Criminal proceeds can be divided among accounts, transferred through intermediaries, recombined, and eventually sent to apparently legitimate destinations. The risk associated with one account may therefore depend partly on its relationships with other actors and on its structural position in the transaction network. Conventional classification methods that describe transactions or accounts independently may fail to capture this information.

Bitcoin provides a useful setting in which to study these issues. Bitcoin transfers are recorded on a public blockchain, making it possible to reconstruct transactions among pseudonymous wallet addresses. At the same time, users can create multiple addresses and transfer funds repeatedly through new addresses, exchanges, mixers, and other cryptocurrency services. These features can generate complex transaction chains that obscure the economic relationship between the original and final holders of funds. Bitcoin has consequently become an important setting for both illicit-finance research and the development of network-based detection methods \citep{weber2019elliptic,bellei2024elliptic2}.

Bitcoin transactions also have an intrinsically higher-order structure. A transaction can contain multiple input addresses and multiple output addresses, so a representation based exclusively on pairwise edges may discard information about addresses jointly participating in the same transaction. Directed graphs describe the direction and connectivity of fund flows, whereas directed hypergraphs can preserve the many-to-many relationship between sets of input and output addresses. These representations may therefore provide structural information beyond conventional address-level variables.

This paper studies whether such structural information improves illicit-address detection and whether it can reveal economically meaningful patterns in suspicious fund-flow chains. We use the Elliptic++ Bitcoin dataset and construct sixteen label-independent graph and directed-hypergraph features, including PageRank, betweenness centrality, weak-component size, transaction-set-size measures, hyperedge size, expansion ratio, and co-participation measures. These features are added to the original Elliptic++ address characteristics and evaluated using Logistic Regression, Random Forest, Multilayer Perceptron, and Extreme Gradient Boosting (XGB) classifiers. The original Elliptic++ features do include ``one-hop'' neighborhood characteristics, but our additional features often consider the network neighborhood beyond one-hop. This is potentially important for anti-money laundering applications because more extensive sub-network flows are likely to be involved. Also, importantly, our inclusion of directed-hypergraph features brings in the joint nature of groupings of transactions and their relationships to other groupings in the blockchain. Considering that money-laundering often involves extensive layering and mixing of flows, this may also be helpful for anti-money laundering work. Thus, our study provides evidence as to whether the relationship of a transaction to the broader network of transaction is additionally useful.

A particularly important feature of the data is that most wallet addresses are unlabeled (i.e., not known by Elliptic, the blockchain analytics firm that produced it, to be either licit or illicit). Simply removing these addresses would eliminate a large part of the transaction network, whereas assuming that unknown addresses are licit would introduce an unjustified classification. We instead retain unknown addresses and use the trained classifier to estimate their illicit-possibilities. 

Our empirical results show that network structure contains useful information for illicit-address detection. Adding graph and directed-hypergraph features improves the performance of several classifiers, with the enriched XGB model providing the strongest overall results. Relative to the model using only the original address variables, structural enrichment substantially improves illicit-class recall and F1-score while preserving high precision. The result suggests that an address's network position and higher-order transaction relationships provide predictive information beyond its isolated characteristics.

This, in turn, makes it possible to connect address-level prediction to a second important question: how difficult or how easy might it be to disrupt money-laundering activities based on the information in the predictions? We call this assessment the ``fragility'' of money laundering. 

To assess fragility, we begin by constructing directed motifs, i.e., sequences of transactions connecting illicit starting transactions with licit ending transactions, on which the fragility of money laundering behaviors can be tested. Illicit-to-licit paths are not uniformly distributed across unknown addresses. When unknown intermediaries with the highest predicted illicit-risk scores are progressively removed, connectivity disappears for a substantial fraction of the corresponding motifs. Removing these connections effectively blocks the sequence of exchanges in that alternative connection paths tend not to exist. The network is therefore relatively fragile with respect to a concentrated set of high-risk intermediaries. This result is potentially important for AML applications because it suggests that predictive models can be combined with network structure to identify not only suspicious actors but also intermediaries whose investigation or removal would affect a disproportionate share of suspicious fund-flow connections.

The contribution of this paper is threefold. First, we introduce label-independent graph and directed-hypergraph variables that extend the original Elliptic++ address features dataset. Second, we evaluate whether these structural variables improve illicit-address classification and use the resulting model to estimate money-laundering risk among previously unlabeled addresses. Third, we further examine the fragility of motifs that are likely to be associated with money-laundering activity and find that targeted disruption of a small number of high-risk motifs can produce a disproportionately large reduction in connectivity. Although our empirical application uses Bitcoin, we regard this is a proof-of-concept framework can also be applied to other banking, payment, trade-credit, securities in which risk depends on relationships among interconnected actors.

The remainder of the paper is organized as follows. Section~2 reviews the related literatures on cryptocurrency illicit-activity detection, graph and hypergraph methods, and financial-network analysis, and also places our analysis within the existing research on illicit financial activity and financial networks. Section~3 describes the problems on which we focus. Section~4 introduces the data structure. Section~5 presents the classification tasks and empirical results. Section~6 develops the money laundering motif and network-fragility analysis. The final section discusses implications, limitations, and conclusions.

\section{Foundations in the Literature}

This section reviews six strands of prior work that jointly motivate the empirical design of this paper. Weber et al. (2019) introduce the original Elliptic dataset and demonstrate the value of graph-based information for illicit-transaction classification. Ranshous et al. (2017) show that Bitcoin transactions can be represented as directed hypergraphs and that recurring transaction motifs may reveal economically meaningful intermediary behavior. Elmougy and Liu (2023) extend this framework through Elliptic++, which adds wallet-address data and multiple network representations that make address-level forensic analysis possible. Bellei et al. (2024) shift the focus from isolated transactions to complete illicit-to-licit subgraph structures, providing a direct motivation for studying suspicious fund-flow motifs. Accominotti et al. (2023) provide a higher-order framework for measuring the resilience and fragility of financial intermediation chains under targeted node removal. Finally, Safar et al. (2026) identify important reproducibility and data-leakage concerns in the original Elliptic feature construction, motivating the leakage-controlled design used in this study. Taken together, these papers establish the three building blocks of our analysis: network-based illicit-risk prediction, higher-order representation of financial flows, and structural fragility analysis. The remainder of this section reviews each contribution in greater detail and explains how it informs the research questions, data design, classification experiments, and motif-removal analysis developed in the subsequent sections.

\citet{weber2019elliptic} introduce the original Elliptic dataset. This dataset is a transaction-level Bitcoin graph. Each node is a Bitcoin transaction, and each directed edge represents Bitcoin flow from one transaction to another. The dataset contains 203,769 transaction nodes and 234,355 directed payment-flow edges. The labels are illicit, licit, and unknown. A small fraction of transactions are labeled illicit, a larger fraction are labeled licit, and the majority are unknown.

The original Elliptic dataset provides 166 features for each transaction \citep{weber2019elliptic}. These include local transaction features and aggregated neighborhood features. Local features describe information about the transaction itself, such as the number of inputs and outputs, transaction fee, output volume, and related transaction statistics. Aggregated features summarize one-hop backward and forward neighborhood information. These aggregated features are important because Bitcoin transaction risk is not only a property of a transaction itself; it is also related to the nearly money-flow neighborhood. We extend the inquiry by considering broader network characteristics and hypergraph relational groupings. 

Using the original Elliptic data, \citet{ranshous2017stb} introduced the idea of the ``short thick band'' (STB) to investigate whether a simple exchange-centered transaction structure could help identify potential Bitcoin money laundering. They modeled the blockchain as a directed hypergraph, in which addresses are vertices and transactions are directed hyperedges connecting multiple input addresses to multiple output addresses. An STB is a two-transaction motif in which Bitcoin moves from an exchange address, passes through one or more intermediate addresses, and then reaches either the same exchange or another exchange; critically, the middle addresses are not labeled as exchanges. 

\citet{ranshous2017stb} first trained several classifiers to distinguish exchange from non-exchange addresses using 19 address-level features. Model 1 included three features derived from known exchange labels and achieved an F1 score of 0.9973, whereas Model 2 removed these label-dependent features and used only 16 structural features, reducing the F1 score to 0.8200. This comparison was necessary because Model 1 relied heavily on incomplete exchange labels and therefore generalized poorly to unlabeled addresses. When applied to STB middle addresses, Model 1 predicted only 1.34\%--0.68\% as exchanges, while Model 2 predicted 48.35\%--52.09\%, compared with 18.17\% of ordinary unlabeled addresses. Thus, the middle addresses of STBs behaved structurally more like exchanges than normal users. Contrary to the original laundering hypothesis, the results suggest that STBs mainly capture hidden exchange-controlled addresses, self-churn, or inter-exchange activity rather than providing direct evidence of money laundering. 

The implication for our analysis is that the network structure of unlabelled transactions alone cannot be depended upon to identify money laundering. Information about the labels of the intermediate transactions will need to be added. Information about wallets, which is not present in the original dataset, would also likely be useful.

\citet{elmougy2023ellipticplusplus} introduce the Elliptic\texttt{++} dataset, which extends the original Elliptic dataset \citep{weber2019elliptic} by adding over 822,000 Bitcoin wallet addresses, each with 56 features, and more than 1.27 million temporal address occurrences or interactions \citep{elmougy2023ellipticplusplus}. Elliptic\texttt{++} also enhances the transaction features \citep{elmougy2023ellipticplusplus}. In addition to the 166 original Elliptic features, it adds 17 augmented transaction features. The authors perform fraud detection on both transactions and wallet addresses \citep{elmougy2023ellipticplusplus}. \citet{elmougy2023ellipticplusplus} show that Random Forest with feature refinement performs strongly for both transaction-level and actor-level detection. This supports the methodological choice to use Elliptic++ for our experiments. Importantly, note that Elliptic++ does not include hypergraph-based characterisitcs as we do. 

\citet{bellei2024elliptic2} bring an explicit graph approach to the analysis, and also introduce Elliptic2, a Bitcoin subgraph dataset designed to represent complete money-laundering patterns rather than isolated suspicious transactions. The study emphasizes that laundering activity is often expressed through directed subgraphs in which funds originate from illicit services, pass through intermediate addresses or transactions, and ultimately reach exchanges or other licit services. This subgraph perspective is directly relevant to the present paper because it supports defining suspicious structures as ordered illicit-to-licit paths instead of treating every edge independently.

The Elliptic2 framework also distinguishes structural evidence from verified criminal attribution. A path from an illicit source through unlabeled intermediaries to a licit destination is a candidate laundering pattern, but its intermediate unknown nodes do not automatically become ground-truth illicit actors. This paper inspires us regarding the designation of the highly-likely money laundering motifs.

Our hypergraph approach is also partly motivated a study of the importance of transactions groupings represented as a hypergraph that is from outside the AML setting. \citet{accominotti2023intermediaries} provides a close methodological analogue to the fragility experiment developed in this paper. Their central insight is that a financial intermediation chain should be treated as a complete higher-order structure rather than as a collection of independent pairwise links. Their hyperstructure approach preserves the internal organization of each chain and allows the importance of an intermediary to be evaluated by whether other participants can still access the network through alternative chains. Using node-removal simulations, they show that network resilience depends not only on how connected an intermediary is, but also on how easily that intermediary can be substituted. This logic is directly applicable to our Bitcoin setting. Their node-removal framework therefore motivates our test of whether targeted removal of high-risk unknown addresses causes suspicious fund-flow motifs to disappear or lose alternative routes.

Although Elliptic++ data have been important for methodological research in AML, an important limitation has recently been noted by \citet{safar2026opacity}. Their reverse engineering of the original Elliptic dataset shows that several features described as local contain address-level or temporally aggregated information, and that preprocessing, look-ahead information, and repeated-address exposure can inflate reported out-of-sample performance. Nevertheless, our analysis provides a proof-of-concept for our approach by comparison to other papers that use the same data, in that we achieve additional results. 

The literature reviewed above provides the necessary foundations for this study, but it also leaves an important gap. Existing work has separately examined directed-hypergraph transaction patterns, graph-based illicit-activity prediction, richer address-level Bitcoin datasets, suspicious subgraph structures, financial-network resilience, and data-leakage concerns. However, these elements have not been integrated into a single framework that first uses leakage-controlled graph and directed-hypergraph features to estimate illicit risk at the wallet-address level and then uses those predictions to test the fragility of potential money-laundering motifs. This paper fills that gap by linking predictive classification to structural intervention: the first empirical component evaluates whether higher-order network information improves illicit-address detection, while the second examines whether the illicit activities on the cryptocurrency network are fragile. In this way, the literature reviewed in this section maps directly onto the empirical architecture of the paper, which is Sections 4 and 5. Building on these foundations, Section 3 now formalizes the two research problems that organize the empirical analysis.

\section{Research Problems}

This section formalizes the two research problems that organize the empirical analysis of the paper. The first establishes whether directed hypergraph information improves illicit-address prediction and generates risk scores for unknown addresses, while the second uses those scores to evaluate the fragility of candidate money-laundering paths. These two questions form a sequential empirical framework: the first produces the risk estimates used to rank unknown intermediaries, while the second uses those estimates to evaluate the disruptability of suspicious fund-flow structures. The following subsections define these two problems and the corresponding empirical objects in detail.

\subsection{Research Problem 1: Structural Feature Enrichment and \\Unknown-Address Risk Prediction}
\label{subsec:first_contribution}

The first research problem asks:

\begin{quote}
\textit{Does augmenting the original Elliptic++ wallet-address feature space with newly mined graph-based and directed-hypergraph-based features improve the predictive performance of machine-learning models for Bitcoin anti-money-laundering classification?}
\end{quote}

This question is evaluated by training Logistic Regression, Random Forest, Multilayer Perceptron, and Extreme Gradient Boosting models on two feature spaces: the original Elliptic++ actor features and a leakage-controlled feature space enriched with label-independent graph and directed-hypergraph variables. Unknown addresses are excluded from supervised training because their classes are unresolved. After the comparative experiment identifies the strongest classifier, that trained model is applied to unknown addresses to estimate a continuous illicit-risk probability. These probabilities are not treated as ground-truth labels; they provide a ranking of unknown intermediaries for the second experiment.

\subsection{Research Problem 2: Fragility of Illicit-to-Licit Motifs}
\label{subsubsec:fragility_illicit_chains}

The next research problem examined in this study concerns the structural
fragility of potential money-laundering chains in the Bitcoin transaction
network. Detecting a suspicious transaction or address is not sufficient to
determine whether the associated illicit flow structure is robust or whether
it depends on a small number of critical intermediaries. We therefore ask:

\begin{quote}
\textit{How fragile are potential money-laundering motifs in the Bitcoin
transaction network?}
\end{quote}

This question is motivated by the hyperstructure approach of Accominotti
et al., in which a financial intermediation chain is treated as a complete
higher-order structure rather than as a collection of independent pairwise
links. Under this interpretation, the removal of one node invalidates every
chain containing that node, including relationships between other nodes that
do not directly involve the removed node. Consequently, a conventional
dyadic graph may underestimate fragility because it does not preserve the
unity and internal organization of the complete financial chain.

The same reasoning applies to Bitcoin money flows. A potential
money-laundering motif may connect an illicit source to an intermediate
address and subsequently to another illicit or licit destination. The
continued existence of this motif depends on all addresses and transactions
that constitute the complete chain. If an intermediate address is removed,
blocked, seized, or otherwise made unavailable, the entire observed
fund-flow path may disappear unless an alternative complete path remains.
Accordingly, a structure is resilient when the suspicious
flow can continue through another valid chain, whereas it is
fragile when the removal of the address disconnects the relevant source and
destination.

The resulting fragility analysis has a direct AML interpretation. A fragile
money-laundering structure may be disrupted effectively by targeting a small
number of critical addresses, whereas a resilient structure may continue to
operate despite the removal of individual participants. Measuring this
difference provides evidence about the concentration, substitutability, and
potential disruptability of suspicious Bitcoin fund-flow chains.

The Elliptic family of datasets was developed specifically to support empirical research on illicit activity in Bitcoin networks. The original Elliptic dataset, introduced by \citet{weber2019elliptic}, represented Bitcoin transactions as a directed payment-flow graph and provided transaction-level features and licit, illicit, and unknown labels for evaluating machine-learning methods in financial forensics. It subsequently became a widely used benchmark for Bitcoin AML research because it offered one of the first large-scale public datasets linking real blockchain transactions with illicit-activity labels. \citet{elmougy2023ellipticplusplus} later extended this framework through Elliptic\texttt{++}, adding wallet-address-level information. This development is particularly important for the present study because our two research problems require both address-level risk prediction and the reconstruction of higher-order fund-flow structures. Elliptic\texttt{++} therefore serves not merely as a convenient data source, but as the central empirical benchmark that makes the classification and motif-fragility analyses in this paper possible.

Together, these two research problems define the empirical logic of the study.  The remainder of the paper implements these questions sequentially: Section~4 describes the data, network representations, and classification models; Section~5 evaluates the first research problem; and Section~6 evaluates the second. We therefore turn next to the Elliptic++ dataset and the modeling framework used to conduct the empirical analysis.

\section{Dataset and Model Description}

This section describes the Elliptic++ data, its transaction- and address-level structure, the relevant labels and features, and the classification models used to implement the two research problems defined above.

\subsection{Elliptic++ Dataset Description}

\subsubsection{Information of Transactional Edges}

The Elliptic++ dataset is the advance version of the Elliptic dataset.
The original Elliptic++ dataset contains two categories of data, the first is the information of transactions and the second is the information of actors, which are the wallet addresses in the bitcoin transactional network. There are 203769 transactions and 234355 directed transaction-to-transaction edges. The definition of transaction-to-transaction edges is as follows: a directed edge from transaction \(v_i\) to transaction \(v_j\) exists if at least one output generated by \(v_i\) is subsequently spent as an input to \(v_j\). Therefore, the direction of the edge follows the flow of Bitcoin value from the source transaction to the destination transaction.

The actor component contains 822,942 unique Bitcoin wallet addresses and 2,868,964 directed address-to-address interaction records. An address-to-address edge from address \(a_i\) to address \(a_j\) exists when \(a_i\) is used as an input address and \(a_j\) is used as an output address in the same Bitcoin transaction. Therefore, the direction of the edge follows the flow of Bitcoin value from the sending address to the receiving address. If the same ordered pair of addresses interacts through multiple transactions, multiple edge records may be present. Among the 2,868,964 address-to-address interaction records, 2,784,344 correspond to unique ordered address pairs, while 84,620 represent repeated interactions between previously observed address pairs.

Elliptic++ also provides a heterogeneous address--transaction graph. It contains 477,117 directed input-address-to-transaction edges and 837,124 directed transaction-to-output-address edges. An input-address-to-transaction edge \(a_i \rightarrow v_j\) indicates that address \(a_i\) supplies Bitcoin as an input to transaction \(v_j\), whereas a transaction-to-output-address edge \(v_j \rightarrow a_k\) indicates that transaction \(v_j\) transfers Bitcoin to output address \(a_k\). Together, these two edge types form 1,314,241 directed address--transaction edges and represent the complete flow of Bitcoin from input addresses, through transactions, to output addresses.

The Elliptic++ dataset preserves the temporal structure of the original Elliptic dataset and contains 49 discrete time steps covering approximately January 2016 to October 2017. Consecutive time steps are separated by approximately two weeks, and each time step contains a connected component of transactions confirmed within a window of less than approximately three hours. Thus, each time step represents an approximately instantaneous snapshot rather than a complete two-week transaction network. No transaction-to-transaction edges connect different time steps, so the transaction dataset consists of 49 temporally ordered but mutually disconnected subgraphs:

\begin{equation}
G_T
=
\bigcup_{t=1}^{49} G_T^{(t)},
\qquad
E_T^{(t_i,t_j)}
=
\varnothing
\quad
\text{for } t_i \neq t_j.
\end{equation}

The uploaded transaction data confirm that all 49 time steps are represented. The number of transactions in an individual time step ranges from 1,089 to 7,880, with a mean of approximately 4,158.55 and a median of 4,291 transactions per time step \cite{weber2019elliptic,bellei2019elliptic}.

In the actor dataset, wallet addresses are assigned to the time steps of the transactions in which they participate. Each address-to-address interaction is generated by a single Bitcoin transaction and therefore belongs to the same time step as that transaction. Thus, actor-to-actor edges do not directly connect addresses across different time steps. However, the same wallet address, or even the same ordered pair of addresses, may reappear in multiple time steps through different transactions. Therefore, the actor network can also be represented as 49 temporally ordered snapshots, although their node sets are not necessarily disjoint \cite{elmougy2023ellipticplusplus}.

This temporal organization also creates an important distribution-shift problem for illicit-activity detection. In particular, time step 43 occurred after the shutdown of a darknet market.

\subsubsection{Information of Labels on Transactions and Wallet Addresses}
\label{subsec:transaction_actor_labels}

Both transactions and actors are assigned one of three class labels:

\begin{equation}
label=
\begin{cases}
1, & \text{illicit},\\
2, & \text{licit},\\
3, & \text{unknown}.
\end{cases}
\end{equation}

\noindent{\textbf{Transaction-label definition and origin}}

In the original Elliptic dataset, a transaction is labelled according to the entity that initiated it. The initiating entity is defined as the entity believed to control the private keys associated with the input addresses of the transaction. A transaction is labelled licit when its initiating entity belongs to a known legitimate category, such as an exchange, wallet provider, miner, financial-service provider, or another licit service. It is labelled illicit when its initiating entity belongs to a known illicit category, such as a scam, malware operation, ransomware operation, Ponzi scheme, terrorist organization, darknet market, or another criminal service \cite{weber2019elliptic,bellei2019elliptic}.

The transaction labels are therefore based on entity attribution rather than solely on the numerical properties of a transaction. The original study further notes that this simplified attribution is less straightforward for mixing transactions because the input addresses of one transaction may be controlled by multiple entities \cite{weber2019elliptic}. Transactions for which Elliptic did not have sufficiently reliable licit or illicit attribution information were labelled unknown. Thus, the unknown category does not represent a separate behavioral type between licit and illicit; instead, it indicates unresolved attribution and may contain both genuinely licit and genuinely illicit transactions.

Among the 203,769 transactions, 4,545 are labelled illicit, 42,019 are labelled licit, and 157,205 are labelled unknown. Their distribution is shown in Table~\ref{tab:transaction_label_distribution}.

Table~\ref{tab:transaction_label_distribution} shows that, among the 46,564 transactions with known binary labels, 4,545 (9.7608\%) are illicit, while 42,019 (90.2392\%) are licit, indicating a substantial class imbalance.
\begin{table}[htbp]
\centering
\caption{Distribution of transaction labels in Elliptic++}
\label{tab:transaction_label_distribution}
\begin{tabular}{lrr}
\toprule
Transaction class & Number & Percentage \\
\midrule
Illicit & 4,545 & 2.2305\% \\
Licit & 42,019 & 20.6209\% \\
Unknown & 157,205 & 77.1486\% \\
\midrule
Total & 203,769 & 100.0000\% \\
\bottomrule
\end{tabular}
\end{table}


\begin{table}[htbp]
\centering
\caption{Distribution among transactions with known binary labels}
\label{tab:labelled_transaction_distribution}
\begin{tabular}{lrr}
\toprule
Transaction class & Number & Percentage \\
\midrule
Illicit & 4,545 & 9.7608\% \\
Licit & 42,019 & 90.2392\% \\
\midrule
Total & 46,564 & 100.0000\% \\
\bottomrule
\end{tabular}
\end{table}

\noindent{\textbf{Actor-label definition and origin}}

The actor (wallets)  dataset contains 822,942 unique Bitcoin wallet addresses. Elliptic++ derives the label of an address from the classes of the transactions with which that address is connected. 

Among the 822,942 unique wallet addresses, 14,266 are labelled illicit, 251,088 are labelled licit, and 557,588 are labelled unknown. Their distribution is shown in Table~\ref{tab:unique_actor_label_distribution}.

\begin{table}[htbp]
\centering
\caption{Distribution of labels among unique wallet addresses}
\label{tab:unique_actor_label_distribution}
\begin{tabular}{lrr}
\toprule
Actor class & Number & Percentage \\
\midrule
Illicit & 14,266 & 1.7335\% \\
Licit & 251,088 & 30.5110\% \\
Unknown & 557,588 & 67.7554\% \\
\midrule
Total & 822,942 & 100.0000\% \\
\bottomrule
\end{tabular}
\end{table}

After unknown addresses are excluded, the binary actor sample contains 265,354 unique addresses. Of these, 14,266, or 5.3763\%, are illicit and 251,088, or 94.6237\%, are licit.

Because the same wallet address may appear in more than one time step, the actor feature dataset contains 1,268,260 address--time observations. These observations consist of 28,601 illicit, 338,871 licit, and 900,788 unknown address occurrences.

\begin{table}[htbp]
\centering
\caption{Distribution of actor labels among address--time observations}
\label{tab:actor_time_label_distribution}
\begin{tabular}{lrr}
\toprule
Actor class & Address--time observations & Percentage \\
\midrule
Illicit & 28,601 & 2.2551\% \\
Licit & 338,871 & 26.7194\% \\
Unknown & 900,788 & 71.0255\% \\
\midrule
Total & 1,268,260 & 100.0000\% \\
\bottomrule
\end{tabular}
\end{table}

Table~\ref{tab:actor_time_label_distribution} shows that unknown observations constitute the majority of the actor feature dataset at 71.0255\%, followed by licit observations at 26.7194\%, whereas illicit observations account for only 2.2551\%.

A detailed listing and discussion of the features of the Elliptic++ is provided in Appendix A.


\subsection{Classification Models and Graph- and Directed-Hypergraph
Feature Enrichment}
\label{subsec:models_and_feature_enrichment}

In this section, we describe the additional graph- and directec hypergraph features we add to the Elliptic++ dataset.

To preserve the temporal structure of Elliptic++, time steps \(1\)--\(34\)
were used for training and time steps \(35\)--\(49\) were used for testing.
The temporal split was performed before feature scaling. The MinMaxScaler
was fitted only on the training set and then applied unchanged to the test
set; therefore, no test-set feature values were used to fit either the scaler
or the classifier. The same procedure was applied to both the original and
enriched actor feature datasets.

Logistic Regression is a linear classification model that estimates the
probability of a binary outcome using the logistic function. It learns a
weighted combination of input features and classifies an address as illicit
or licit according to the resulting probability
\citep{berkson1944logistic}.

Random Forest is an ensemble learning method that combines multiple
decision trees trained on randomly sampled data and features. Each tree
produces a class prediction, and the final classification is determined by
majority voting, which improves robustness and captures nonlinear feature
interactions \citep{breiman2001randomforest}.

A Multilayer Perceptron is a feedforward neural network consisting of
interconnected layers of neurons. It learns nonlinear relationships among
input features through nonlinear activation functions and adjusts network
weights using backpropagation to minimize classification error
\citep{rumelhart1986learning}.

The fourth classifier is a gradient-boosted decision-tree model. It builds
decision trees sequentially, with each new tree focusing on correcting errors
made by the existing ensemble, while regularization controls model
complexity. The predictions of all trees are combined to produce the final
classification \citep{chen2016xgboost}.

The formal definitions of the directed graph and directed hypergraph used
throughout this paper are provided in
Appendix~\ref{para:mining_address_graph_hypergraph_features}. Building on
those definitions, one of the main contributions of this study is the
construction of 16 additional graph- and directed-hypergraph-based actor
features that enrich the original Elliptic++ actor feature dataset. These
features describe structural information that is not explicitly represented
by the original behavioral actor variables, including network importance,
intermediary position, proximity to labelled actors, participation in
higher-order transactions, transaction expansion and consolidation, and
co-participation with other wallet addresses.

Let the label of address \(a\) be

\[
    y_a\in\{1,2,3\},
\]

\noindent where \(y_a=1\), \(y_a=2\), and \(y_a=3\) indicate illicit, licit, and
unknown addresses, respectively. Based on the directed graph and directed
hypergraph representations defined in Appendix~\ref{para:mining_address_graph_hypergraph_features}, the following 16
graph-based and directed-hypergraph-based address features are constructed.

\begin{description}

\item[\textbf{1. \texttt{addr\_pagerank}.}]
For a given wallet address \(a\), its PageRank score is denoted by
\[
    \pi_a.
\]
The score \(\pi_a\) measures the importance of address \(a\) in the directed
address graph.

To define this feature clearly, let

\[
    u
\]
denote an address that sends funds directly to \(a\), so that

\[
    (u,a)\in E_{\mathrm{addr}}.
\]

Let

\[
    v
\]
denote any address to which \(u\) sends funds. Therefore, the out-degree of
\(u\) is

\[
    d_u^{\mathrm{out}}
    =
    \left|
        \left\{
            v\in V:
            (u,v)\in E_{\mathrm{addr}}
        \right\}
    \right|,
\]
which is the number of distinct addresses receiving directed connections
from \(u\).

The PageRank score of address \(a\) is then defined as

\[
    \pi_a
    =
    \frac{1-\alpha}{|V|}
    +
    \alpha
    \sum_{\substack{u\in V\\(u,a)\in E_{\mathrm{addr}}}}
    \frac{\pi_u}{d_u^{\mathrm{out}}},
    \qquad
    \alpha=0.85.
\]

In this formula, \(\pi_u\) is the PageRank score of an address \(u\) that
points to \(a\). The quantity

\[
    \frac{\pi_u}{d_u^{\mathrm{out}}}
\]

is the portion of \(u\)'s PageRank score passed to address \(a\). If \(u\)
points to many different addresses, its score is divided among those
addresses. If \(u\) points to only a few addresses, each receiving address
obtains a larger portion of \(\pi_u\).

Thus, address \(a\) obtains a high PageRank score when it receives directed
connections from important addresses, especially when those addresses have
only a small number of outgoing connections. The term

\[
    \frac{1-\alpha}{|V|}
\]

assigns every address a small baseline score, while the second term
aggregates the importance transferred to \(a\) from all addresses \(u\)
that directly point to it.

\item[\textbf{2. \texttt{addr\_betweenness}.}]
For each address \(a\), this feature measures how often \(a\) acts as an
intermediate node on shortest directed fund-flow paths.

The source set \(\mathcal{S}\) contains 64 wallet addresses selected from
all addresses with positive out-degree. These addresses are first ordered
by out-degree, and 64 positions are selected at approximately equal
intervals across that ordering to form the set \(\mathcal{S}\). For each sampled source address
\(s\in\mathcal{S}\), every address \(t\) reachable from \(s\) through the
directed address graph is treated as a possible destination.

Let \(\sigma_{st}\) be the number of shortest directed paths from \(s\) to
\(t\), and let \(\sigma_{st}(a)\) be the number of those paths that pass
through \(a\). \(V\) is the set of all wallet addresses represented as nodes in the directed address graph. The approximate betweenness centrality is
\[
    \widetilde{C}_{B}(a)
    =
    \frac{|V|}
    {|\mathcal{S}|(|V|-1)(|V|-2)}
    \sum_{s\in\mathcal{S}}
    \sum_{\substack{t\in V\\t\neq s,\;t\neq a}}
    \frac{\sigma_{st}(a)}{\sigma_{st}}.
\]
Here, \(s\) is one of the 64 sampled starting addresses, \(t\) is a
destination address reachable from \(s\), and \(a\) is the address whose
feature is being calculated. A larger approximate betweenness centrality value means that \(a\) more
frequently lies between other addresses on their shortest directed
fund-flow paths. 

\item[\textbf{3. \texttt{addr\_weak\_component\_size}.}]
This feature measures the size of the connected address group containing
address \(a\).

To compute it, the directions of all edges are temporarily ignored. Two
addresses are treated as connected if one can be reached from the other
through a sequence of address-to-address links, regardless of the original
direction of fund flow.

Let
\[
    \operatorname{WCC}(a)
\]
denote the set of all addresses belonging to the same weakly connected
component as address \(a\). The feature is defined as
\[
    W_a
    =
    \left|
        \operatorname{WCC}(a)
    \right|,
\]
where \(\left|\operatorname{WCC}(a)\right|\) is the number of addresses in
that component. A larger value means that address \(a\) belongs to a larger
connected group of wallet addresses. As shown in Fig.~\ref{fig:weak_component_size}, a larger weakly connected component size indicates that address \(a\) belongs to a larger connected wallet cluster.

\begin{figure}[htbp]
    \centering
    \includegraphics[width=0.42\textwidth]{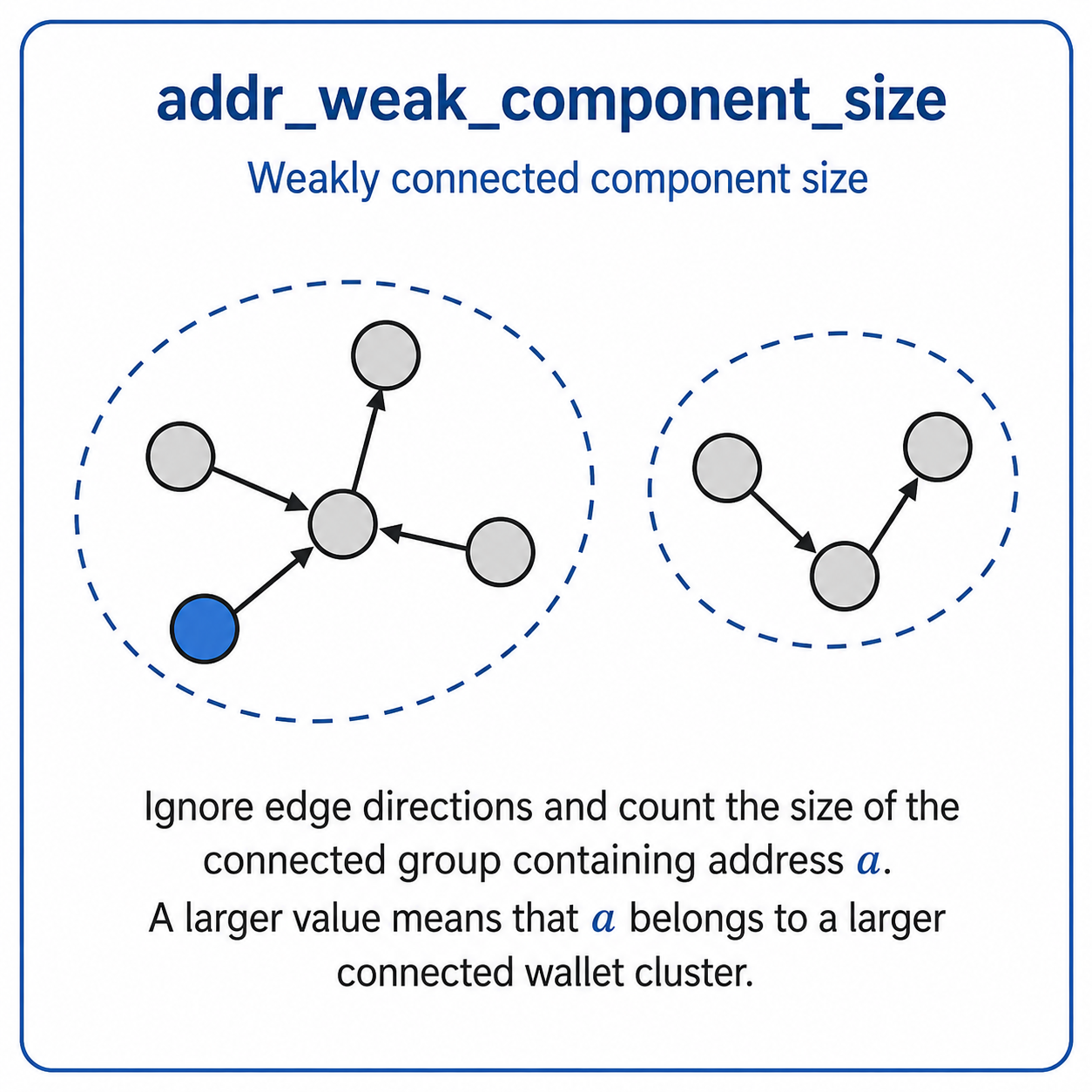}
    \caption{Illustration of the weakly connected component size of an address.}
    \label{fig:weak_component_size}
\end{figure}

\item[\textbf{4. \texttt{addr\_illicit\_neighbor\_count}.}]
The number of distinct illicit one-hop neighbors of address \(a\) is
\[
    N_a^{I}
    =
    \left|
        \left\{
            u\in\mathcal{N}(a):
            y_u=1
        \right\}
    \right|.
\]

\item[\textbf{5. \texttt{addr\_illicit\_neighbor\_ratio}.}]
The proportion of the one-hop neighbors of address \(a\) that are labelled
illicit is
\[
    R_a^{I}
    =
    \begin{cases}
        \displaystyle
        \frac{N_a^{I}}
        {|\mathcal{N}(a)|},
        &
        |\mathcal{N}(a)|>0,\\[10pt]
        0,
        &
        |\mathcal{N}(a)|=0.
    \end{cases}
\]

\item[\textbf{6. \texttt{addr\_unknown\_neighbor\_ratio}.}]
The proportion of the one-hop neighbors of address \(a\) that are labelled
unknown is
\[
    R_a^{U}
    =
    \begin{cases}
        \displaystyle
        \frac{
            \left|
                \left\{
                    u\in\mathcal{N}(a):
                    y_u=3
                \right\}
            \right|
        }
        {|\mathcal{N}(a)|},
        &
        |\mathcal{N}(a)|>0,\\[12pt]
        0,
        &
        |\mathcal{N}(a)|=0.
    \end{cases}
\]

\item[\textbf{7. \texttt{addr\_distance\_to\_illicit\_min}.}]
Let
\[
    \operatorname{dist}_{G^{(u)}}(a,u)
\]
denote the length of the shortest unweighted path between addresses \(a\)
and \(u\) in the symmetrized graph. The minimum distance from address \(a\)
to any known illicit address is
\[
    D_a^{I}
    =
    \min_{u:y_u=1}
    \operatorname{dist}_{G^{(u)}}(a,u).
\]
If no illicit address is reachable from \(a\), the feature is encoded as
\[
    D_a^{I}=-1.
\]

\item[\textbf{8. \texttt{addr\_num\_illicit\_within\_2hop}.}]
The number of distinct illicit addresses located within at most two
undirected hops of address \(a\) is
\[
    K_a^{I,2}
    =
    \left|
        \left\{
            u\in V\setminus\{a\}:
            y_u=1,\,
            \operatorname{dist}_{G^{(u)}}(a,u)\leq 2
        \right\}
    \right|.
\]

\item[\textbf{9. \texttt{addr\_avg\_input\_set\_size}.}]
For address \(a\), this feature is the average number of input addresses in
all transactions involving \(a\):
\[
\overline{n}_a^{\mathrm{in}}
=
\begin{cases}
\displaystyle
\frac{1}{h_a}
\sum_{e_t\in\mathcal{E}(a)}
|T_t|,
&
h_a>0,\\[8pt]
0,
&
h_a=0.
\end{cases}
\]
Here, \(\mathcal{E}(a)\) is the set of transactions containing \(a\),
\(h_a=|\mathcal{E}(a)|\) is the number of those transactions, and
\(|T_t|\) is the number of distinct input addresses in transaction \(t\).

\item[\textbf{10. \texttt{addr\_avg\_output\_set\_size}.}]
For address \(a\), this feature is the average number of output addresses in
all transactions involving \(a\):
\[
\overline{n}_a^{\mathrm{out}}
=
\begin{cases}
\displaystyle
\frac{1}{h_a}
\sum_{e_t\in\mathcal{E}(a)}
|H_t|,
&
h_a>0,\\[8pt]
0,
&
h_a=0.
\end{cases}
\]
Here, \(\mathcal{E}(a)\) is the set of transactions containing \(a\),
\(h_a=|\mathcal{E}(a)|\) is the number of those transactions, and
\(|H_t|\) is the number of distinct output addresses in transaction \(t\).

\item[\textbf{11. \texttt{addr\_max\_hyperedge\_size}.}]
For address \(a\), this feature is the largest hyperedge cardinality among
all transactions containing \(a\):
\[
M_a
=
\begin{cases}
\displaystyle
\max_{e_t\in\mathcal{E}(a)}
|e_t|,
\\[8pt]
0,
\end{cases}
\qquad
\]
Here, \(\mathcal{E}(a)\) is the set of transactions containing \(a\),
\(h_a=|\mathcal{E}(a)|\), and \(|e_t|\) is the hyperedge cardinality,
namely, the total number of distinct input and output addresses in
transaction \(t\).

\item[\textbf{12. \texttt{addr\_avg\_expansion\_ratio}.}]
For address \(a\), this feature is the average ratio of the number of
distinct output addresses to the number of distinct input addresses across
all transactions containing \(a\). A value greater than \(1\) indicates
that these transactions tend to distribute funds to more output addresses,
whereas a value below \(1\) indicates that they tend to consolidate funds
from more input addresses into fewer output addresses.

\item[\textbf{13. \texttt{addr\_co\_input\_degree}.}]
For address \(a\), this feature counts how many other input addresses appear
together with \(a\) in the same transactions:
\[
C_a^{\mathrm{in}}
=
\sum_{t:\,a\in T_t}
\left(|T_t|-1\right).
\]
Here, \(T_t\) is the set of input addresses in transaction \(t\), and
\(|T_t|-1\) excludes \(a\) itself. A larger value means that \(a\) more
often appears together with other addresses on the input side of
transactions.

\item[\textbf{14. \texttt{addr\_co\_output\_degree}.}]
For address \(a\), this feature counts how many other output addresses appear
together with \(a\) in the same transactions:
\[
C_a^{\mathrm{out}}
=
\sum_{t:\,a\in H_t}
\left(|H_t|-1\right).
\]
Here, \(H_t\) is the set of output addresses in transaction \(t\), and
\(|H_t|-1\) excludes \(a\) itself. A larger value means that \(a\) more
often appears together with other addresses on the output side of
transactions.

\item[\textbf{15. \texttt{addr\_illicit\_hyperedge\_ratio}.}]
For address \(a\), this feature is the proportion of its transactions that
contain at least one illicit address:
\[
R_a^{I,e}
=
\frac{\text{number of transactions containing \(a\) and at least one illicit address}}
{\text{total number of transactions containing \(a\)}}.
\]
A larger value means that address \(a\) more often participates in
transactions involving known illicit addresses.

\item[\textbf{16. \texttt{addr\_unknown\_bridge\_count}.}]
This feature is calculated only for an unknown address \(a\). It counts how
many transactions containing \(a\) also contain at least one illicit address
and at least one other address:
\[
B_a^{U}
=
\#\left\{
t:
a\in T_t\cup H_t,\;
y_a=3,\;
\exists\,u\in T_t\cup H_t \text{ with } y_u=1,\;
|T_t\cup H_t|\geq 3
\right\}.
\]
Here, \(t\) is a transaction, \(T_t\) and \(H_t\) are its input and output
address sets, and \(y_a=3\) means that address \(a\) is labelled unknown.
A larger value means that \(a\) more often appears in mixed transactions
together with illicit and other addresses.

\end{description}

These 16 variables form the complete set of newly constructed graph- and
directed-hypergraph-based actor features and constitute an extension of the
original Elliptic++ actor feature dataset. The four classification models are evaluated first on the original
Elliptic++ actor feature space and then on the enriched,
leakage-controlled actor feature space. Their classification performance is
compared in Section~5. The model that achieves the strongest overall
performance on the enriched feature space is subsequently selected for
estimating illicit risk among the unknown addresses. This selected model,
rather than a model chosen in advance, provides the risk probabilities used
in the second experiment.

\subsection{Prediction of Illicit Risk for Unknown Addresses Using the
Best-Performing Enriched-Feature Model}
\label{subsec:unknown_address_prediction}

The next stage connects the first classification experiment with the
motif-fragility experiment. The model that performs best in the
classification task using the original Elliptic++ actor features augmented
by the newly constructed graph- and directed-hypergraph feature set is
subsequently applied to all wallet addresses labelled as unknown in the
original Elliptic++ dataset.

The purpose of this procedure is not to replace the original labels, but to
assign each unknown address a model-estimated probability of being illicit.

The known addresses are encoded as

\[
y_a=
\begin{cases}
1, & \text{if address } a \text{ is illicit},\\
0, & \text{if address } a \text{ is licit}.
\end{cases}
\]

Let

\[
x_a
=
[x_{a1},x_{a2},\ldots,x_{an}]^{\top}
\]

\noindent denote the leakage-controlled feature vector for address
\(a\).

The best-performing model is trained on the known-label training matrix and
evaluated on the held-out known-label test matrix. It is then applied
without retraining to the feature matrix of the 557,588 unique unknown
addresses. For each unknown address \(a\), the selected model produces

\[
p_a
=
\Pr(y_a=1\mid x_a),
\]

\noindent where \(x_a\) is the feature vector of address \(a\). The
probability of being licit is calculated as

\[
1-p_a.
\]

\noindent Then a provisional binary prediction is produced using a threshold of
\(0.5\):

\[
\widehat{y}_a
=
\begin{cases}
1, & p_a\geq0.5,\\
2, & p_a<0.5.
\end{cases}
\]

\noindent Here, the value \(1\) denotes a predicted illicit address and the value
\(2\) denotes a predicted licit address, following the original Elliptic++
label convention. These predicted labels remain model outputs rather than
verified ground-truth labels.

The distribution of predicted illicit probabilities is reported in the
empirical-results section rather than used here to identify the selected
classifier. At this stage, the methodological role of the prediction is
simply to transform the classification results from Experiment~1 into an
ordered risk measure for unknown intermediary addresses. Experiment~2 then
uses this ordering to evaluate whether removing the most highly ranked
unknown intermediaries disproportionately reduces or eliminates
illicit-to-licit fund-flow connectivity.

\section{Experiment 1: Classification with Graph- and \\ Hypergraph-Enriched Features}

This section tests the first research problem by comparing illicit-address classification performance before and after graph- and directed-hypergraph features are added to the original Elliptic++ actor feature space.

\subsection{Set-Up}

In the first set of experiments, we add the newly-added graph-based and hypergraph-based features to the feature vectors of the actor dataset of Elliptic++. We separately train four different models on the original Elliptic++ address feature dataset and the Graph and Hypergraph based enhanced Elliptic++ address feature dataset, aiming at getting 8 models that are able to classify the labels one and two on the addresses. Then we compare the experimental results to find which set of models is better at the  classification task, and judge whether the new graph-based and hypergraph-based features improve the predictive performance of artificial-intelligence models.

\subsubsection{Rationale for Using Wallet Addresses as the Modeling Unit.}
The classification models in this study are trained on wallet addresses rather than transactions because the address-level data provide a more appropriate setting for evaluating the incremental contribution of the newly constructed graph and directed-hypergraph features. First, the original Elliptic transaction feature set does not provide a suitable control-variable design for this comparison. Each transaction is represented by 166 features, of which the first 94 are local features describing characteristics such as the number of input and output addresses, transaction fees, transferred volumes, and statistical properties of the associated inputs and outputs. The remaining 72 features are generated by aggregating these local attributes over one-hop neighbouring transactions using statistics such as the minimum, maximum, standard deviation, and correlation coefficient \citep{weber2019elliptic}. Consequently, the complete transaction feature set already mixes intrinsic transaction attributes with neighbourhood-derived information. If the newly constructed graph or hypergraph features were added directly to this transaction feature set, there would be no clearly defined control group containing only non-structural baseline features. It would therefore be difficult to determine whether any observed improvement resulted from the proposed graph and hypergraph representations or from graph-related information that was already embedded in the 72 aggregated transaction features.

In contrast, the original Elliptic++ address features are explicitly defined in terms of the transaction histories and financial behaviours of individual wallet addresses. Their meanings can therefore be examined feature by feature, making it possible to verify that the selected baseline address features do not already contain the explicit graph- and hypergraph-topological quantities introduced in this study.

A second reason concerns the semantic origin of the class labels. Although the nodes in the original Elliptic graph represent transactions, Weber et al.\ state that a transaction is labelled licit or illicit according to the real-world entity that initiates it, namely, the entity controlling the private keys associated with the transaction's input addresses \citep{weber2019elliptic}. Thus, the label assigned to a transaction is not determined exclusively by the transaction itself. Instead, it reflects the licit or illicit status of the entity controlling the relevant wallet addresses. Transaction labels can therefore be interpreted as transaction-level manifestations of an underlying address- or entity-level status. Wallet addresses provide a more direct analytical unit for learning this underlying behaviour, because they represent the persistent entities through which multiple transactions are initiated, received, and connected over time. For these reasons, address-level modeling is both methodologically necessary for establishing an appropriate control-variable framework and conceptually better aligned with the entity-based meaning of the Elliptic labels.

\subsection{Results}
\label{subsubsec:actor_classification_comparison}

Four classifiers, namely Logistic Regression (LR), Random Forest (RF),
Multilayer Perceptron (MLP), and Extreme Gradient Boosting (XGB), were
first trained on the original Elliptic++ actor feature set to distinguish
illicit addresses from licit addresses. Addresses labelled unknown were
excluded, and the remaining labels were recoded as
\[
y_i =
\begin{cases}
1, & \text{if address } i \text{ is illicit},\\
0, & \text{if address } i \text{ is licit}.
\end{cases}
\]
The same four classifiers were subsequently trained on the enriched
actor feature set. Although the enriched dataset contained the 16
graph- and directed-hypergraph feature candidates defined in this
study, seven label-informed features were excluded to prevent target
leakage. The leakage-controlled models therefore used the 55 original
numerical actor features together with nine newly constructed,
label-independent structural features. Specifically, the seven label-informed features excluded from model
training were \texttt{addr\_illicit\_neighbor\_count},
\texttt{addr\_illicit\_neighbor\_ratio},
\texttt{addr\_unknown\_neighbor\_ratio},
\texttt{addr\_distance\_to\_illicit\_min},
\texttt{addr\_num\_illicit\_within\_2hop},
\texttt{addr\_illicit\_hyperedge\_ratio}, and
\texttt{addr\_unknown\_bridge\_count}.

Precision, recall, and F1 score were calculated with the illicit class
treated as the positive class:
\[
P_{\mathrm{ill}}
=
\frac{TP}{TP+FP},
\qquad
R_{\mathrm{ill}}
=
\frac{TP}{TP+FN},
\]
and
\[
F1_{\mathrm{ill}}
=
\frac{2P_{\mathrm{ill}}R_{\mathrm{ill}}}
     {P_{\mathrm{ill}}+R_{\mathrm{ill}}}.
\]
This choice reflects the practical objective of anti-money-laundering
detection, for which illicit-class recall is particularly important
because a false negative corresponds to an illicit address that remains
undetected. Accordingly, the precision, recall, and F1 values reported
in Tables~\ref{tab:original_actor_results} and
\ref{tab:enriched_actor_results} quantify classification performance
for illicit addresses rather than for licit addresses.

For \(C\) classes, in this case \(C\) = 2, micro-averaged precision and recall are defined as
\[
P_{\mathrm{micro}}
=
\frac{\sum_{c=1}^{C}TP_c}
     {\sum_{c=1}^{C}(TP_c+FP_c)},
\qquad
R_{\mathrm{micro}}
=
\frac{\sum_{c=1}^{C}TP_c}
     {\sum_{c=1}^{C}(TP_c+FN_c)},
\]
and Micro-F1 is
\[
F1_{\mathrm{micro}}
=
\frac{2P_{\mathrm{micro}}R_{\mathrm{micro}}}
     {P_{\mathrm{micro}}+R_{\mathrm{micro}}}.
\]
For a single-label binary classification problem,
\[
F1_{\mathrm{micro}}
=
\frac{TP+TN}{TP+TN+FP+FN},
\]
which is equal to overall classification accuracy.

\begin{table}[htbp]
\centering
\caption{Illicit-address classification results using the original
Elliptic++ address features}
\label{tab:original_actor_results}
\begin{tabular}{lcccc}
\toprule
Model & Precision & Recall & F1 Score & Micro-F1 \\
\midrule
LR  & 0.477 & 0.046 & 0.083 & 0.964 \\
RF  & \textbf{0.911} & \textbf{0.789} &
      \textbf{0.845} & \textbf{0.990} \\
MLP & 0.708 & 0.502 & 0.587 & 0.974 \\
XGB & 0.869 & 0.534 & 0.662 & 0.980 \\
\bottomrule
\end{tabular}

\vspace{1mm}
\begin{minipage}{0.94\linewidth}
\footnotesize
\textit{Note:} The classifiers were trained using the original
address-level features provided in the Elliptic++ dataset. Precision,
recall, and F1 score are reported for the illicit-address class,
whereas Micro-F1 evaluates overall classification performance.
\end{minipage}
\end{table}

As shown in Table~\ref{tab:original_actor_results}, RF achieved the
strongest performance on the original actor feature set. It obtained
an illicit-class precision of \(0.911\), recall of \(0.789\), and F1
score of \(0.845\). Its Micro-F1 was \(0.990\). LR produced the weakest
illicit-address detection result because its recall was only \(0.046\),
indicating that it failed to identify most illicit addresses.

\begin{table}[htbp]
\centering
\caption{Illicit-address classification results using the
leakage-controlled graph- and directed-hypergraph-enriched
Elliptic++ address features}
\label{tab:enriched_actor_results}
\begin{tabular}{lcccc}
\toprule
Model & Precision & Recall & F1 Score & Micro-F1 \\
\midrule
LR  & 0.708 & 0.064 & 0.118 & 0.965 \\
RF  & \textbf{0.945} & 0.823 & 0.880 & 0.992 \\
MLP & 0.819 & 0.480 & 0.605 & 0.977 \\
XGB & 0.938 & \textbf{0.841} &
      \textbf{0.887} & \textbf{0.992} \\
\bottomrule
\end{tabular}

\vspace{1mm}
\begin{minipage}{0.94\linewidth}
\footnotesize
\textit{Note:} The enriched dataset contained 16 graph- and
directed-hypergraph feature candidates. Seven label-informed features
were excluded to prevent target leakage, leaving nine additional
structural features for model training. Precision, recall, and F1 score
are reported for the illicit-address class.
\end{minipage}
\end{table}

As shown in Table~\ref{tab:enriched_actor_results}, XGB achieved the
best overall illicit-address classification performance after the
graph- and directed-hypergraph features were incorporated. It attained
a precision of \(0.938\), recall of \(0.841\), F1 score of \(0.887\),
and Micro-F1 of \(0.992\). RF achieved the highest precision,
\(0.945\), but its recall and F1 score were slightly lower than those
of XGB. Thus, RF generated marginally fewer false-positive illicit
alerts, whereas XGB identified a larger proportion of the genuinely
illicit addresses and provided the strongest balance between precision
and recall.

For AML applications, illicit-class recall is operationally important
because a false negative corresponds to an illicit address that is not
detected. Nevertheless, maximizing recall alone may generate an
unacceptably large number of false-positive alerts. Accordingly,
illicit-class F1 score is used as the principal overall comparison
metric, while recall is examined as the principal missed-illicit-risk
metric. Micro-F1 is not used as the primary model-selection criterion
because the large number of licit addresses causes it to be dominated
by majority-class predictions.

As for the prediction of illicit risk for unknown addresses using the
best-performing enriched-feature modelEach row of the resulting dataset corresponds to exactly one unknown Bitcoin
address. The address identifier was preserved during preprocessing, so the
predicted probability is directly matched to the same address. The resulting
CSV contains the address identifier, original class, predicted licit probability,
predicted illicit probability, provisional predicted label, descending illicit-risk
rank, and illicit-probability percentile. The descending ranking is used in the
subsequent fragility experiment to remove the unknown addresses with the highest
predicted illicit probabilities first.

Table~\ref{tab:unknown_xgb_probability_distribution} summarizes the distribution
of predicted illicit probabilities among the 557,588 unknown addresses.

\begin{table}[htbp]
\centering
\caption{Distribution of XGB-predicted illicit probabilities among unknown addresses}
\label{tab:unknown_xgb_probability_distribution}
\begin{tabular}{lrr}
\toprule
Predicted illicit-probability interval &
Number of addresses &
Percentage \\
\midrule
$[0.00,0.10)$ & 486,067 & 87.1731\% \\
$[0.10,0.25)$ & 31,319 & 5.6169\% \\
$[0.25,0.50)$ & 18,761 & 3.3647\% \\
$[0.50,0.75)$ & 9,350 & 1.6769\% \\
$[0.75,0.90)$ & 4,810 & 0.8626\% \\
$[0.90,1.00]$ & 7,281 & 1.3058\% \\
\midrule
Total & 557,588 & 100.0000\% \\
\bottomrule
\end{tabular}
\end{table}

The mean predicted illicit probability was 0.0587, whereas the median was only
0.0009. At the conventional \(0.5\) threshold, 21,441 unknown addresses
(3.8453\%) were predicted to be illicit. In addition, 7,281 addresses
(1.3058\%) received an illicit probability of at least \(0.90\), and 5,374
addresses (0.9638\%) received a probability of at least \(0.95\). These results
indicate that the model assigns low illicit probabilities to most unknown
addresses, while identifying a comparatively small high-risk subset. The
continuous probabilities, rather than only the thresholded labels, are therefore
retained for the subsequent progressive node-removal and motif-fragility analysis.

\subsection{Interpretation of the Experiment and Results}

The results show that adding graph- and directed-hypergraph-based features generally improves illicit-address classification. The original Elliptic++ actor features mainly describe the local transaction behavior of each wallet address, whereas the newly added features capture its structural position, intermediary role, and participation in multi-address transactions.

XGB benefited most from the graph- and hypergraph-enriched features because its boosting mechanism is well suited to learning complex interactions among tabular structural variables. LR assumes a mainly linear relationship between each feature and illicit risk, so it cannot easily capture cases in which a graph feature becomes informative only when combined with other features. RF can model nonlinear relationships, but its trees are trained independently and then averaged. In contrast, XGB builds trees sequentially, with each new tree focusing on the classification errors left by the previous trees. This allows it to progressively exploit subtle combinations of centrality, hyperedge size, co-input degree, co-output degree, and expansion ratio. MLP can also learn nonlinear relationships, but it may be less stable on highly imbalanced tabular data without additional tuning or class-balancing methods. Therefore, XGB extracted the greatest predictive benefit from the newly added structural features, increasing illicit-class precision from 0.869 to 0.938, recall from 0.534 to 0.841, and F1 score from 0.662 to 0.887.

The hypergraph features are particularly useful because Bitcoin transactions are often many-to-many structures involving multiple input and output addresses. Potential laundering behavior may include consolidating funds from several addresses, distributing funds to many outputs, or repeatedly co-spending with other addresses. Ordinary pairwise graphs do not fully preserve this joint participation structure. Directed-hypergraph features therefore provide additional information about fund aggregation, dispersion, and coordinated address behavior.

Overall, the improved results indicate that illicit activity is associated not only with the individual financial behavior of an address, but also with its position and role in the wider transaction structure. Combining ordinary graph features with directed-hypergraph features produces a more complete representation of Bitcoin fund flows and substantially improves the detection performance of XGB. As a direction for future research, an XGB feature-contribution table could be constructed to quantify the relative importance of the original and newly added graph- and directed-hypergraph features in driving the model's predictions.

\section{Experiment 2: Motif Fragility under Targeted Node Removal}

This section tests the second research problem by progressively removing intermediaries from high-risk money laundering motifs and measuring how strongly the resulting fund-flow connectivity deteriorates or disappears.

\subsection{Set-Up}

\subsubsection{Operational Research Questions for the Fragility Experiment}

The fragility experiment answers two operational sub-questions. First, \textbf{Does the motif break after the targeted removal of predicted high-risk unknown intermediaries?} Second, \textbf{How many unknown intermediaries must be removed before source-to-destination connectivity is destroyed, and how much of the original path set is lost at each removal stage?} These questions separate the binary event of path failure from the degree of connectivity degradation.

\subsubsection{Motif Definitions and Structural Interpretation}
\label{par:potential_money_laundering_motifs}

Unless otherwise stated, all graph- and directed-hypergraph notation used in this subsection follows the definitions in Appendix~\ref{para:mining_address_graph_hypergraph_features}; in particular, \(V\) denotes the set of wallet addresses, \(\mathcal{E}\) the set of directed transaction hyperedges, and \(e_t=(T_t,H_t)\) a transaction hyperedge with input-address set \(T_t\) and output-address set \(H_t\).

Using the directed transaction hypergraph
\(\mathcal{H}=(V,\mathcal{E})\) defined above, let
\[
a_i \rightsquigarrow_{\mathcal{H}} a_j
\]
denote a directed address-to-address transfer induced by at least one
transaction hyperedge \(e_t=(T_t,H_t)\in\mathcal{E}\), such that
\[
a_i\in T_t
\qquad\text{and}\qquad
a_j\in H_t.
\]
Thus, \(a_i\rightsquigarrow_{\mathcal{H}}a_j\) means that address \(a_i\)
appears on the input side and address \(a_j\) appears on the output side
of the same directed transaction hyperedge.

For \(k\in\{1,2,3\}\), an illicit-to-licit motif with \(k\) unknown
intermediaries is defined as the ordered address sequence
\[
\mathcal{P}_k
=
\left(
a_0
\rightsquigarrow_{\mathcal{H}}
a_1
\rightsquigarrow_{\mathcal{H}}
\cdots
\rightsquigarrow_{\mathcal{H}}
a_k
\rightsquigarrow_{\mathcal{H}}
a_{k+1}
\right),
\]
where
\[
y_{a_0}=1,\qquad
y_{a_i}=3\ \text{for }1\leq i\leq k,
\qquad
y_{a_{k+1}}=2,
\]
and all addresses in the sequence are distinct. Therefore, the three
motif structures considered in this study are
\[
\begin{aligned}
\mathcal{P}_1 &:\
\text{Illicit}
\rightsquigarrow_{\mathcal{H}}
\text{Unknown}
\rightsquigarrow_{\mathcal{H}}
\text{Licit},\\
\mathcal{P}_2 &:\
\text{Illicit}
\rightsquigarrow_{\mathcal{H}}
\text{Unknown}
\rightsquigarrow_{\mathcal{H}}
\text{Unknown}
\rightsquigarrow_{\mathcal{H}}
\text{Licit},\\
\mathcal{P}_3 &:\
\text{Illicit}
\rightsquigarrow_{\mathcal{H}}
\text{Unknown}
\rightsquigarrow_{\mathcal{H}}
\text{Unknown}
\rightsquigarrow_{\mathcal{H}}
\text{Unknown}
\rightsquigarrow_{\mathcal{H}}
\text{Licit}.
\end{aligned}
\]

These motifs follow the suspicious-path interpretation of
\citet{bellei2024elliptic2}, in which funds originating from an illicit
source ultimately reach a licit cluster or legitimate cryptocurrency
service. The intermediate unknown addresses may perform layering, pass-through,
or obfuscation functions before the funds reach an exchange, wallet
provider, or another licit service. The motifs are therefore interpreted
as structural indicators of potential money-laundering activity rather
than as conclusive evidence of criminal conduct
\citep{bellei2024elliptic2}.

After duplicate induced address transfers and self-loops were removed,
each motif was counted once for every distinct ordered address tuple. The
number of motifs with \(k\) unknown intermediaries is
\[
N_k
=
\left|
\left\{
(a_0,\ldots,a_{k+1})\in V^{k+2}:
\begin{array}{l}
a_i\neq a_j\ \text{for all }i\neq j,\\
a_j\rightsquigarrow_{\mathcal{H}}a_{j+1},
\ 0\leq j\leq k,\\
y_{a_0}=1,\ 
y_{a_i}=3\ (1\leq i\leq k),\
y_{a_{k+1}}=2
\end{array}
\right\}
\right|,
\qquad k\in\{1,2,3\}.
\]

\begin{table}[htbp]
\centering
\caption{Frequencies of illicit-to-licit motifs in the Elliptic++
directed transaction hypergraph}
\label{tab:illicit_to_licit_motif_frequencies}

\begin{tabularx}{\textwidth}{
    >{\raggedright\arraybackslash}X
    >{\centering\arraybackslash}p{3.2cm}
    >{\centering\arraybackslash}p{2.3cm}
}
\toprule
Motif structure
&
Number of unknown intermediaries
&
Number of motifs
\\
\midrule

Illicit $\rightsquigarrow_{\mathcal{H}}$ Unknown
$\rightsquigarrow_{\mathcal{H}}$ Licit
&
1
&
5,331
\\

Illicit $\rightsquigarrow_{\mathcal{H}}$ Unknown
$\rightsquigarrow_{\mathcal{H}}$ Unknown
$\rightsquigarrow_{\mathcal{H}}$ Licit
&
2
&
20,452
\\

Illicit $\rightsquigarrow_{\mathcal{H}}$ Unknown
$\rightsquigarrow_{\mathcal{H}}$ Unknown
$\rightsquigarrow_{\mathcal{H}}$ Unknown
$\rightsquigarrow_{\mathcal{H}}$ Licit
&
3
&
57,509
\\

\bottomrule
\end{tabularx}

\vspace{2pt}
\begin{minipage}{\textwidth}
\footnotesize
\textit{Note:} The relation
\(a_i\rightsquigarrow_{\mathcal{H}}a_j\) indicates that \(a_i\) belongs
to the tail set and \(a_j\) belongs to the head set of at least one
directed transaction hyperedge. Duplicate induced address transfers and
self-loops were removed, and each motif was counted once for every
distinct ordered address tuple.
\end{minipage}
\end{table}

As shown in Table~\ref{tab:illicit_to_licit_motif_frequencies}, the
Elliptic++ directed transaction hypergraph contains \(5{,}331\) motifs
with one unknown intermediary, \(20{,}452\) motifs with two unknown
intermediaries, and \(57{,}509\) motifs with three unknown
intermediaries.
\subsubsection{Connectivity and Fragility Measures}
\label{par:connectivity_fragility}

The definitions of directed graphs, directed hypergraphs, and transaction
hyperedges are provided in
Appendix~\ref{para:mining_address_graph_hypergraph_features}. Here, we only
define connectivity and fragility for the node-removal experiment.

In a directed graph, \(v\) is reachable from \(u\) if there exists a directed
path
\[
u=v_0\rightarrow v_1\rightarrow\cdots\rightarrow v_{\ell}=v.
\]
Connectivity in the directed hypergraph is defined analogously: \(v\) is
reachable from \(u\) if there exists an observed sequence of directed
hyperedges connecting \(u\) to \(v\).

A motif is defined as \emph{fragile} if removing a targeted intermediary
eliminates all directed paths from the illicit source to the licit
destination. It is \emph{not fragile} if at least one alternative path
between the same source and destination remains.

\begin{figure}[tbp]
    \centering
    \subfloat[Fragile structure.]{%
        \includegraphics[width=.48\textwidth]{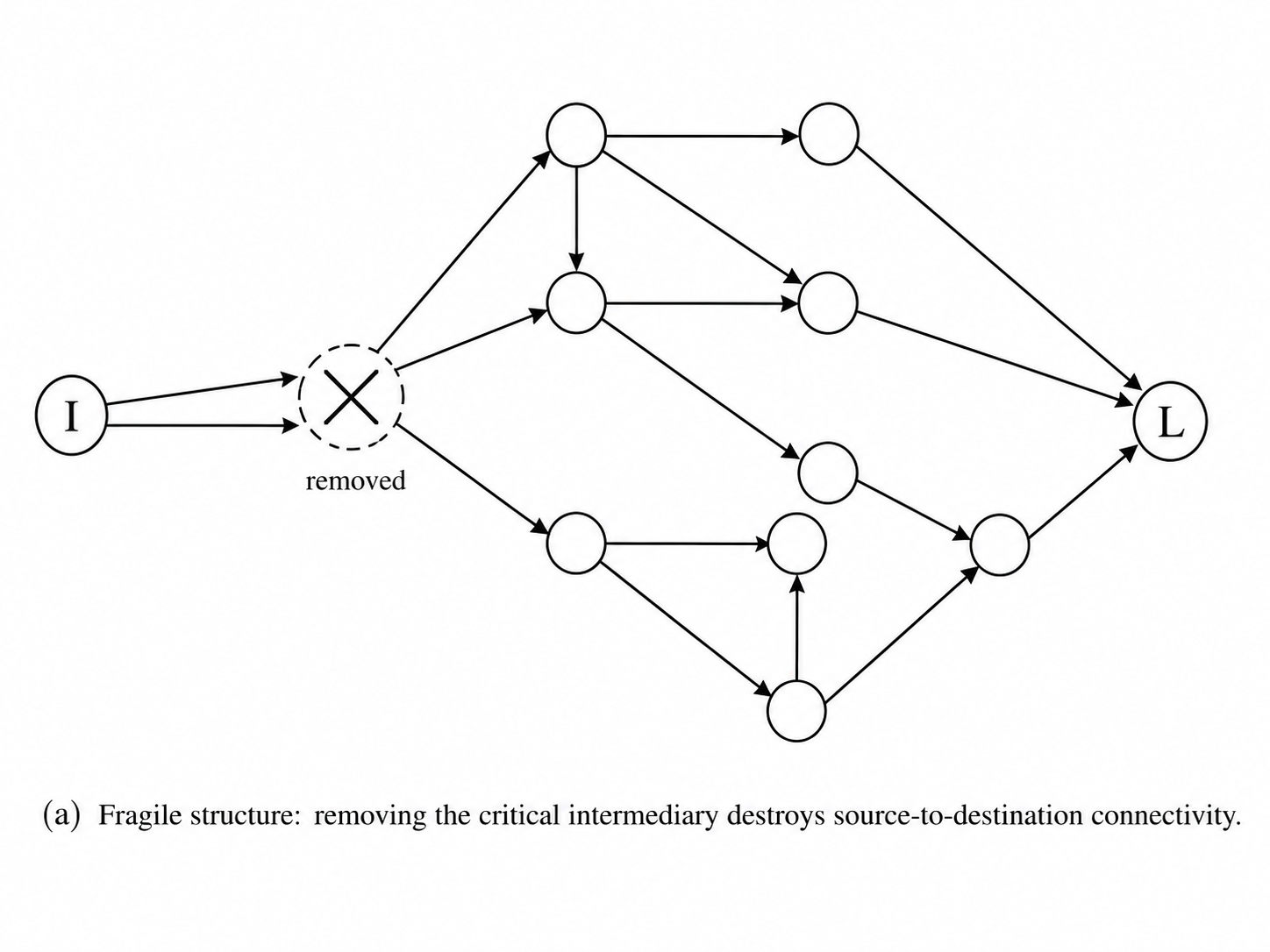}%
    }
    \hfill
    \subfloat[Not-fragile structure.]{%
        \includegraphics[width=.48\textwidth]{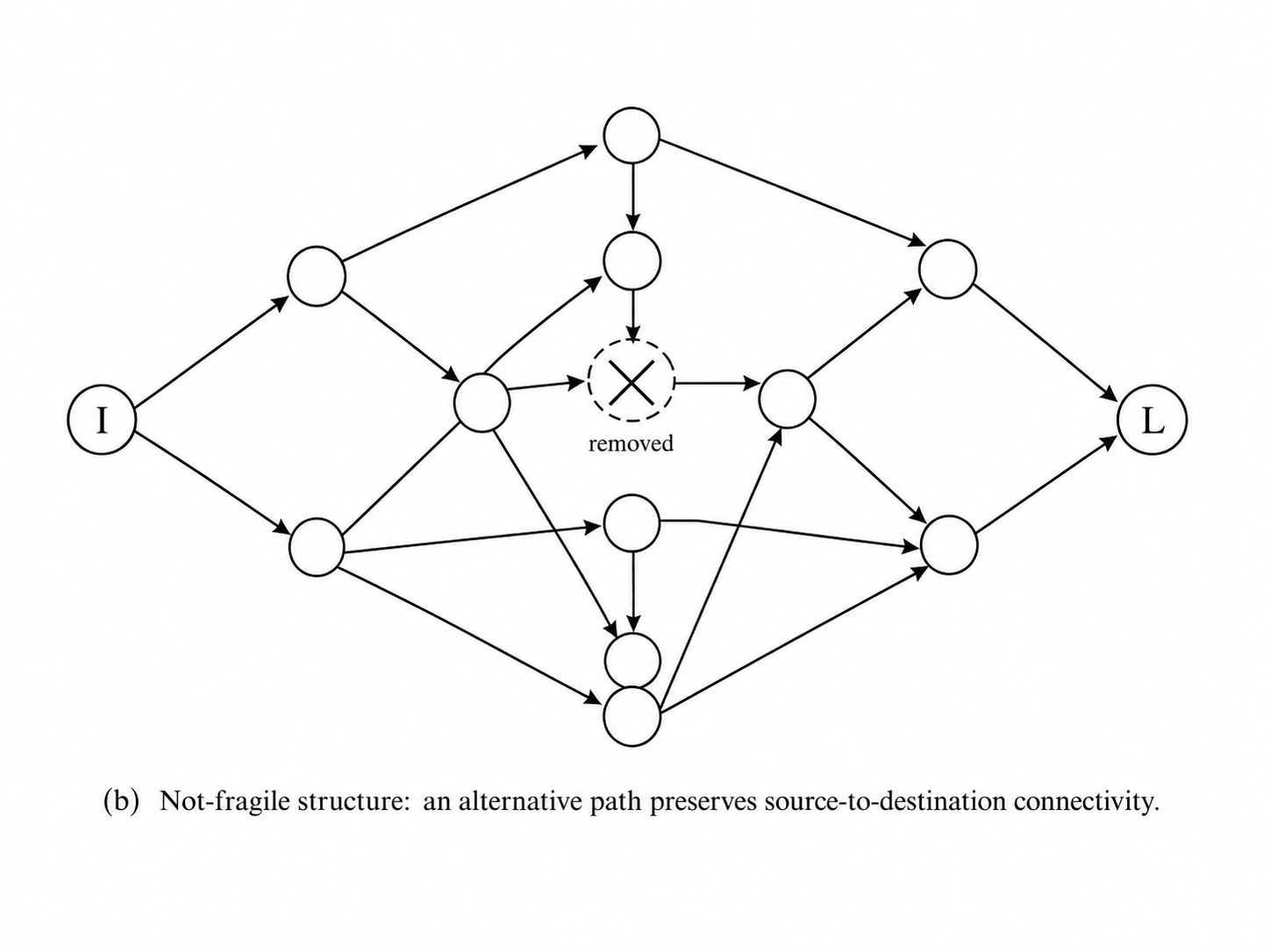}%
    }
    \caption{Fragile vs. not-fragile motif structure.}
    \label{fig:fragile_not_fragile}
\end{figure}

Figure~\ref{fig:fragile_not_fragile} illustrates the distinction. In the
fragile case, all source-to-destination paths depend on the removed
intermediary, whereas in the not-fragile case, other observed paths preserve connectivity. I represents illicit labeled address, and L represents licit labeled address.

\subsubsection{Progressive Removal of Predicted High-Risk Unknown Nodes}

Following the node-removal shock simulation framework of
Accominotti et al.~\cite{accominotti2023intermediaries}, this experiment
evaluates whether the connectivity between the initial illicit address and
the terminal licit address depends on unknown intermediary addresses predicted
to be illicit. For each motif, the unknown intermediary nodes are ranked in
descending order according to their illicit probabilities estimated by the
XGB model. Nodes with the highest predicted probabilities are removed first.
Whenever an intermediary node is removed, all motif instances and directed
paths containing that node are removed before connectivity is recalculated.
No new or hypothetical links are introduced after node removal.

Let \(s_m\) and \(t_m\) denote the initial illicit address and terminal licit
address of motif \(m\), respectively. Let \(P_m^{(0)}\) be the number of
directed paths from \(s_m\) to \(t_m\) before removal, and let
\(P_m^{(r)}\) be the number of remaining paths after removing a proportion
\(r\) of the unknown intermediary nodes. The path-loss ratio is defined as

\[
L_m^{(r)}
=
\frac{P_m^{(0)}-P_m^{(r)}}{P_m^{(0)}}.
\]

The source and terminal addresses remain connected after removal if

\[
P_m^{(r)}>0,
\]

\noindent and connectivity failure occurs if

\[
P_m^{(r)}=0.
\]

For Motif~1, which contains one unknown intermediary node, the single unknown
node is removed. The experiment records whether the removal completely
disconnects the initial illicit address from the terminal licit address.
Because this node represents \(100\%\) of the unknown intermediaries in the
motif, its removal directly tests whether the motif is entirely dependent on
that predicted-illicit intermediary.

For Motif~2, which contains two unknown intermediary nodes, the experiment is
conducted at removal levels of \(50\%\) and \(100\%\). At the \(50\%\) level,
the unknown node with the higher XGB-predicted illicit probability is removed.
At the \(100\%\) level, both unknown nodes are removed. The analysis answers
two questions: first, whether connectivity failure occurs after removing
\(50\%\) or only after removing \(100\%\) of the unknown nodes; and second,
how much the number of directed paths decreases at each removal level.
Accordingly, \(L_m^{(0.50)}\) and \(L_m^{(1.00)}\) are calculated for every
Motif~2 instance.

For Motif~3, which contains three unknown intermediary nodes, the removal
levels are \(33\%\), \(67\%\), and \(100\%\), corresponding to the removal of
one, two, and three unknown nodes, respectively. At each stage, nodes are
removed in descending order of their XGB-predicted illicit probabilities.
The experiment determines whether connectivity failure first occurs after
removing \(33\%\), \(67\%\), or \(100\%\) of the unknown intermediaries.
The corresponding path-loss ratios,
\(L_m^{(0.33)}\), \(L_m^{(0.67)}\), and \(L_m^{(1.00)}\), are also computed.

For each motif type and removal level, the reported statistics include the
number and percentage of motifs that remain connected, the number and
percentage that become disconnected, the mean and median number of remaining
paths, and the mean and median path-loss ratio. This design therefore identifies both the removal
threshold at which motif connectivity fails and the extent to which
connectivity deteriorates before complete disconnection.

\subsubsection{Address Removal and the Same-Owner Limitation}

The removal unit in this experiment is the wallet address. In practice, several addresses may be controlled by the same owner, and removing only one address may not remove the owner's other addresses. Elliptic++ contains user-entity information produced through address-clustering heuristics, but it does not provide complete verified ground-truth ownership for every unknown address. Therefore, the main analysis removes addresses rather than making an unsupported claim that all addresses belonging to the same real-world owner have been identified. Where a reliable user-entity cluster is available, an additional owner-cluster robustness check may remove all addresses assigned to that cluster; however, such a result must be interpreted as heuristic entity-level evidence rather than verified ownership.

\subsection{Results}
\label{subsubsec:motif_fragility_results}

The progressive node-removal experiment described in
Section~3.2.2.3 was applied to the three illicit-to-licit motif structures.
For each motif instance, its unknown intermediary addresses were ranked in
descending order according to their XGB-predicted illicit probabilities. At
each removal stage, the highest-ranked remaining intermediary addresses were
removed, together with all observed motif paths containing those addresses.

An alternative path was defined as another observed path of the same motif
type having the same initial illicit address and the same terminal licit
address. Therefore, connectivity was preserved only when at least one such
alternative motif path remained after node removal. No new or hypothetical
edges were introduced.

For clarity, the four evaluation quantities reported in
Table~\ref{tab:node_removal_fragility_results} are defined as follows.
For a given motif type, let \(N\) denote the total number of motif instances,
and let \(P_m^{(r)}\) denote the number of valid directed paths between the
initial illicit address and terminal licit address of motif instance \(m\)
after removal stage \(r\).

\begin{table*}[htbp]
\centering

\small
\setlength{\tabcolsep}{5pt}
\renewcommand{\arraystretch}{1.10}

\resizebox{\textwidth}{!}{
\begin{tabular}{cccccc}
\toprule
Motif &
Removal level &
\multicolumn{1}{c}{Connected} &
\multicolumn{1}{c}{Disconnected} &
Mean remaining &
Mean path-loss \\
&
&
No. (\%) &
No. (\%) &
paths &
ratio \\
\midrule

$P_1$ &
$100\%$ &
90 (1.6882\%) &
5,241 (98.3118\%) &
0.0195 &
0.9912
\\

$P_2$ &
$50\%$ &
2,786 (13.6221\%) &
17,666 (86.3779\%) &
0.1944 &
0.9308
\\

$P_2$ &
$100\%$ &
39 (0.1907\%) &
20,413 (99.8093\%) &
0.0495 &
0.9986
\\

$P_3$ &
$33\%$ &
5,687 (9.8889\%) &
51,822 (90.1111\%) &
0.7034 &
0.9356
\\

$P_3$ &
$67\%$ &
1,284 (2.2327\%) &
56,225 (97.7673\%) &
0.3641 &
0.9885
\\

$P_3$ &
$100\%$ &
844 (1.4676\%) &
56,665 (98.5324\%) &
0.3217 &
0.9932
\\

\bottomrule
\end{tabular}
}

\caption{Results of the progressive node-removal-based motif fragility analysis. 
Connected is the number (percentage) of endpoint pairs retaining at least one path; 
Disconnected is the number (percentage) retaining no path; 
Mean remaining paths is the average number of surviving paths per endpoint pair; 
Mean path-loss ratio is the average fraction of original paths removed.}
\label{tab:node_removal_fragility_results}

\end{table*}

\subsubsection{Motif 1.}

Motif 1 contains only one unknown intermediary. Removing this address
therefore corresponds to removing 100\% of the unknown intermediary set.
Among the 5,331 Motif 1 instances, 5,241 motifs, or 98.3118\%, became
disconnected after the intermediary was removed. Only 90 motifs, or
1.6882\%, retained at least one alternative path between the same illicit
source and licit destination.

The mean path-loss ratio was 0.9912. These results show that
Motif 1 was almost entirely dependent on its single unknown intermediary.
Thus, the intermediary generally represented a single point of structural
failure.

\subsubsection{Motif 2.}

For Motif 2, the unknown intermediary with the higher XGB-predicted illicit
probability was removed at the 50\% stage. This removal disconnected 17,666
of the 20,452 motifs, corresponding to 86.3779\%. The mean path-loss ratio
was 0.9308, and the mean number of remaining paths was 0.1944.

After both unknown intermediary addresses were removed, 20,413 motifs, or
99.8093\%, were disconnected. The mean path-loss ratio increased to 0.9986.
Only 39 motifs retained an alternative path after both of their original
intermediaries had been removed.

Regarding the first connectivity-failure threshold, 17,666 motifs failed
after removing only the higher-risk intermediary, whereas 2,747 motifs
survived the first removal but failed after the second intermediary was also
removed. The remaining 39 motifs did not fail at either stage. Although most Motif 2 instances
were already dependent on the highest-risk intermediary, removing the second
intermediary produced an additional and statistically significant loss of
connectivity.

\subsubsection{Motif 3.}
For Motif 3, the three unknown intermediaries were removed progressively at
the 33\%, 67\%, and 100\% stages. Removing only the highest-risk
intermediary disconnected 51,822 of the 57,509 motifs, corresponding to
90.1111\%. The mean path-loss ratio at this stage was 0.9356.

After the second-ranked intermediary was also removed, the number of
disconnected motifs increased to 56,225, or 97.7673\%, and the mean
path-loss ratio increased to 0.9885. After all three unknown
intermediaries were removed, 56,665 motifs, or 98.5324\%, were
disconnected, with a mean path-loss ratio of 0.9932.

The first-failure results show that 51,822 motifs failed after the first
removal, 4,403 failed after the second removal, and 440 first failed
only after all three intermediaries were removed. A further 844 motifs
retained at least one alternative path even after all three intermediaries
from the focal motif instance had been removed.

These findings indicate that most Motif 3 instances were concentrated around
one particularly high-risk intermediary. Nevertheless, a smaller subset
required the removal of two or three unknown addresses before connectivity
failed, demonstrating a greater degree of intermediary substitutability.

\section{Conclusion}
\label{subsec:conclusion_future_directions}

In this paper, we have provided a proof-of-concept evaluation of the use of directed-hypergraph features to enhance the detectability of money laundering flows in a blockchain. We have also applied those predictions to investigate the fragility of illicit motifs or sub-chain patterns in the Bitcoin blockchain, such as those that involve money laundering. We find that these motifs are surprisingly fragile. That is, removing wallets or actors that are predicted to be illicit with high probability can sharply impede flows from illicit entry points on the chain to licit exit points.

As we show, the location of wallet accounts in the cryptocurrency graph is useful in providing extra information beyond the original transaction and wallet characteristics available in our Bitcoin dataset. The new graph- and directed-hypergraph-enhanced features clearly improve the identification and ranking of suspicious wallet addresses. The core dataset we use, Elliptic++, has been important in anti-money laundering research. We enhance it by incorporating higher-order, implicit, and topological information that can improve the empirical performance of machine-learning models. Our enhanced data set may, therefore, also be useful to other researchers.  

We believe the performance improvement is because hypergraph-based representations are useful to capture relationships that cannot be fully described by ordinary pairwise networks. In particular, directed hypergraphs provide a more flexible way to represent complex systems in which multiple entities jointly interact and where the direction of flow or influence matters. Researchers can thereby allow the model to learn not only from the attributes of individual nodes, but also from the higher-order organization of the network itself. 

Our fragility findings have encouraging implications for anti-money laundering practice. By breaking small structures within a longer and more hidden flow of funds, it may be possible to destroy a large amount of the original source-to-destination connectivity. This indicates that money laundering could be disrupted by identifying and exclusion of owners of a limited number of crypto wallets.

The framework developed here is not limited to Bitcoin. The underlying research question applies whenever illicit or risky financial activity takes place through interconnected actors and multi-party transactions. Similar methods could be applied to banking payment networks, securities transactions, trade-credit networks, cross-border payments, or other cryptocurrency systems. In each setting, the same general sequence can be investigated: enrich individual-level information with network structure, estimate risk among incompletely labeled actors, identify suspicious higher-order paths, and test whether those paths depend on a small number of structurally critical nodes. This could be a useful heuristic approach to combat illicit activities in modern transactions networks.

\newpage

\appendix{\noindent{\Large\textbf{{Appendices}}}}

\section{Features of Transactions and Wallet Addresses in Elliptic++}

The Elliptic++ dataset provides two numerical feature spaces: a
transaction-level feature space and a wallet-address-level feature space. There are 183 transaction features and 56 actor features in the Elliptic ++ dataset.

\subsection{Transaction features.}

Each of the 203,769 transaction nodes is associated with 183 features:

\begin{equation}
183
=
1+93+72+17.
\end{equation}

These consist of one time-step variable, 93 local transaction features,
72 aggregated neighborhood features inherited from the original
Elliptic dataset, and 17 additional features introduced by
Elliptic++.

\begin{table}[htbp]
\centering
\caption{Composition of the 183 transaction features}
\label{tab:transaction_feature_groups}
\begin{tabular}{p{4.5cm}rp{7.0cm}}
\toprule
Feature group & Number & Interpretation \\
\midrule
Time-step feature
& 1
& Discrete temporal index from 1 to 49 indicating the snapshot in
which the transaction appears. \\

Local transaction features
& 93
& Numerical characteristics of the focal transaction and the historical
behavior of its associated input and output addresses. \\

Aggregated neighborhood features
& 72
& Statistical summaries of transaction characteristics observed in
the one-hop backward and forward neighborhoods of the focal
transaction. \\

Elliptic++ augmented features
& 17
& Additional non-anonymized blockchain variables describing incoming
and outgoing BTC, transaction counts, address counts, fees, and
transaction size. \\
\midrule
Total
& 183
& Complete transaction-level feature vector. \\
\bottomrule
\end{tabular}
\end{table}

\begin{itemize}[
    leftmargin=1.8em,
    label=\textbullet,
    itemsep=1em,
    topsep=0.5em
]

\item \textbf{The 93 local transaction features.}

The original Elliptic dataset provides 93 local numerical features in
addition to the time-step variable. In the released data, these features
are anonymized and are represented as

\[
\mathrm{LF}_1,\mathrm{LF}_2,\ldots,\mathrm{LF}_{93}.
\]

The paper does not disclose a complete one-to-one mapping from every
local-feature index to its underlying variable. It instead explains
that the 93 variables describe transaction-level information including:

\begin{itemize}[
    leftmargin=1.8em,
    label=--,
    itemsep=0.2em,
    topsep=0.3em
]
    \item the number of transaction inputs;
    \item transaction fees;
\end{itemize}
and so on. However, we do not know exactly which underlying feature each $\mathrm{LF}_i$ corresponds to.

\item \textbf{The 72 aggregated neighborhood features.}
The 72 aggregated neighborhood features are graph-derived variables represented as
\[
\mathrm{AF}_1,\mathrm{AF}_2,\ldots,\mathrm{AF}_{72}.
\]
They summarize selected characteristics of transactions located one hop backward or one hop forward from the focal transaction. For example, an aggregated feature may represent the minimum, maximum, standard deviation, or correlation of transaction fees among the focal transaction's neighboring transactions. However, the public dataset does not provide a complete semantic mapping between each index and its underlying feature. Therefore, although these variables are denoted by \(\mathrm{AF}_i\), we do not know exactly which specific aggregated feature each \(\mathrm{AF}_i\) corresponds to.

\item \textbf{The 17 Elliptic++ augmented transaction features.}

Compared to the Elliptic Dataset, Elliptic++ adds 17 explicitly named features obtained from the
deanonymized transaction and address information. The augmented features are listed in
Table~\ref{tab:augmented_transaction_features}.

\begin{table}[htbp]
\centering
\caption{The 17 augmented transaction features in Elliptic++}
\label{tab:augmented_transaction_features}
\begin{tabular}{p{4.0cm}rp{7.8cm}}
\toprule
Feature or feature family & Number & Meaning \\
\midrule
\(\mathrm{BTC}_{\mathrm{in,total}}\)
& 1
& Total amount of BTC entering the transaction through its inputs. \\

\(\mathrm{BTC}_{\mathrm{in,min}}\)
& 1
& Minimum incoming BTC amount among the transaction's incoming
relationships. \\

\(\mathrm{BTC}_{\mathrm{in,max}}\)
& 1
& Maximum incoming BTC amount among the transaction's incoming
relationships. \\

\(\mathrm{BTC}_{\mathrm{in,mean}}\)
& 1
& Mean incoming BTC amount. \\

\(\mathrm{BTC}_{\mathrm{in,median}}\)
& 1
& Median incoming BTC amount. \\

\(\mathrm{BTC}_{\mathrm{out,total}}\)
& 1
& Total amount of BTC transferred from the transaction to its outputs. \\

\(\mathrm{BTC}_{\mathrm{out,min}}\)
& 1
& Minimum outgoing BTC amount among the transaction's outgoing
relationships. \\

\(\mathrm{BTC}_{\mathrm{out,max}}\)
& 1
& Maximum outgoing BTC amount among the transaction's outgoing
relationships. \\

\(\mathrm{BTC}_{\mathrm{out,mean}}\)
& 1
& Mean outgoing BTC amount. \\

\(\mathrm{BTC}_{\mathrm{out,median}}\)
& 1
& Median outgoing BTC amount. \\

\(\mathrm{Txs}_{\mathrm{in}}\)
& 1
& Number of preceding transactions supplying inputs to the focal
transaction. \\

\(\mathrm{Txs}_{\mathrm{out}}\)
& 1
& Number of later transactions that spend outputs of the focal
transaction. \\

\(\mathrm{Addr}_{\mathrm{in}}\)
& 1
& Number of input wallet addresses associated with the transaction. \\

\(\mathrm{Addr}_{\mathrm{out}}\)
& 1
& Number of output wallet addresses associated with the transaction. \\

\(\mathrm{BTC}_{\mathrm{total}}\)
& 1
& Total BTC value transacted. \\

\(\mathrm{Fees}\)
& 1
& Total transaction fees measured in BTC. \\

\(\mathrm{Size}\)
& 1
& Total serialized transaction size on the Bitcoin blockchain. \\
\midrule
Total
& 17
& Complete Elliptic++ augmentation of the original 166 transaction
features. \\
\bottomrule
\end{tabular}
\end{table}

\end{itemize}

\subsection{Actor features.}

Each actor has 56
features. These comprise the time-step variable and 55 numerical
behavioral variables.

The actor features consist of the following categories:

\begin{itemize}[
    leftmargin=1.8em,
    label=--,
    itemsep=0.3em,
    topsep=0.3em
]
    \item one time-step variable;
    \item ten single-valued transaction, lifetime, and block-height
    variables;
    \item fifteen BTC-value variables;
    \item ten fee-related variables;
    \item fifteen block-interval variables; and
    \item five address-interaction summary variables.
\end{itemize}

The feature groups are summarized in
Table~\ref{tab:actor_feature_groups}. The specific definitions of all 56 actor features are provided in
Table~\ref{tab:all_actor_features}.

\begin{table}[htbp]
\centering
\caption{Composition of the 56 actor features}
\label{tab:actor_feature_groups}
\begin{tabular}{p{5.0cm}rp{6.5cm}}
\toprule
Feature group & Number & Interpretation \\
\midrule
Time-step index
& 1
& Temporal snapshot in which the address observation appears. \\

Single-valued transaction and lifetime variables
& 10
& Counts, blockchain heights, lifetime, and temporal persistence of
the address. \\

BTC transacted, sent, and received
& 15
& Five summary statistics for each of three BTC-value families. \\

Fees and fees as a share of BTC
& 10
& Five summary statistics for fees and five for relative fees. \\

Inter-transaction block intervals
& 15
& Five summary statistics for all, input-side, and output-side temporal
intervals. \\

Address-interaction variables
& 5
& Summary statistics describing the address's interactions with
counterparty addresses. \\
\midrule
Total
& 56
& Complete actor-level feature vector. \\
\bottomrule
\end{tabular}
\end{table}

\begingroup
\footnotesize
\setlength{\tabcolsep}{4pt}
\renewcommand{\arraystretch}{1.15}

\FloatBarrier
\begin{longtable}{
    >{\centering\arraybackslash}p{0.06\textwidth}
    >{\RaggedRight\arraybackslash}p{0.34\textwidth}
    >{\RaggedRight\arraybackslash}p{0.52\textwidth}
}
\caption{Definitions and ordering of the 56 actor features}
\label{tab:all_actor_features}\\

\toprule
\textbf{No.} & \textbf{Feature} & \textbf{Meaning} \\
\midrule
\endfirsthead

\multicolumn{3}{c}{
\tablename\ \thetable{} -- continued from previous page
}\\
\toprule
\textbf{No.} & \textbf{Feature} & \textbf{Meaning} \\
\midrule
\endhead

\midrule
\multicolumn{3}{r}{Continued on next page}\\
\endfoot

\bottomrule
\endlastfoot

1
&
\feature{Time step}
&
Discrete temporal index from 1 to 49 for the address--time
observation.
\\

2
&
\feature{num_txs_as_sender}
&
Number of dataset transactions in which the address appears as an
input or sending address.
\\

3
&
\feature{num_txs_as_receiver}
&
Number of dataset transactions in which the address appears as an
output or receiving address.
\\

4
&
\feature{first_block_appeared_in}
&
Bitcoin block height at which the address first appears in the
collected blockchain history.
\\

5
&
\feature{last_block_appeared_in}
&
Bitcoin block height at which the address most recently appears in
the collected blockchain history.
\\

6
&
\feature{lifetime_in_blocks}
&
Difference between the last and first observed block heights,
measuring the observed lifetime of the address in blocks.
\\

7
&
\feature{total_txs}
&
Total number of blockchain transactions involving the address.
\\

8
&
\feature{first_sent_block}
&
Block height at which the address is first observed sending BTC.
\\

9
&
\feature{first_received_block}
&
Block height at which the address is first observed receiving BTC.
\\

10
&
\feature{num_timesteps_appeared_in}
&
Number of Elliptic time steps in which the address appears.
\\

11
&
\feature{btc_transacted_total}
&
Total BTC transacted by the address, including both sent and
received BTC.
\\

12
&
\feature{btc_transacted_min}
&
Minimum BTC amount among transactions involving the address.
\\

13
&
\feature{btc_transacted_max}
&
Maximum BTC amount among transactions involving the address.
\\

14
&
\feature{btc_transacted_mean}
&
Mean BTC amount among transactions involving the address.
\\

15
&
\feature{btc_transacted_median}
&
Median BTC amount among transactions involving the address.
\\

16
&
\feature{btc_sent_total}
&
Total BTC sent by the address.
\\

17
&
\feature{btc_sent_min}
&
Minimum BTC amount sent by the address.
\\

18
&
\feature{btc_sent_max}
&
Maximum BTC amount sent by the address.
\\

19
&
\feature{btc_sent_mean}
&
Mean BTC amount sent by the address.
\\

20
&
\feature{btc_sent_median}
&
Median BTC amount sent by the address.
\\

21
&
\feature{btc_received_total}
&
Total BTC received by the address.
\\

22
&
\feature{btc_received_min}
&
Minimum BTC amount received by the address.
\\

23
&
\feature{btc_received_max}
&
Maximum BTC amount received by the address.
\\

24
&
\feature{btc_received_mean}
&
Mean BTC amount received by the address.
\\

25
&
\feature{btc_received_median}
&
Median BTC amount received by the address.
\\

26
&
\feature{fees_total}
&
Total transaction fees associated with transactions involving the
address.
\\

27
&
\feature{fees_min}
&
Minimum associated transaction fee.
\\

28
&
\feature{fees_max}
&
Maximum associated transaction fee.
\\

29
&
\feature{fees_mean}
&
Mean associated transaction fee.
\\

30
&
\feature{fees_median}
&
Median associated transaction fee.
\\

31
&
\feature{fees_as_share_total}
&
Aggregate transaction fees expressed relative to the amount of BTC
transacted.
\\

32
&
\feature{fees_as_share_min}
&
Minimum fee-to-BTC-transacted ratio.
\\

33
&
\feature{fees_as_share_max}
&
Maximum fee-to-BTC-transacted ratio.
\\

34
&
\feature{fees_as_share_mean}
&
Mean fee-to-BTC-transacted ratio.
\\

35
&
\feature{fees_as_share_median}
&
Median fee-to-BTC-transacted ratio.
\\

36
&
\feature{blocks_btwn_txs_total}
&
Total accumulated block intervals between consecutive transactions
involving the address.
\\

37
&
\feature{blocks_btwn_txs_min}
&
Minimum number of blocks between two consecutive transactions
involving the address.
\\

38
&
\feature{blocks_btwn_txs_max}
&
Maximum number of blocks between two consecutive transactions
involving the address.
\\

39
&
\feature{blocks_btwn_txs_mean}
&
Mean number of blocks between consecutive transactions involving
the address.
\\

40
&
\feature{blocks_btwn_txs_median}
&
Median number of blocks between consecutive transactions involving
the address.
\\

41
&
\feature{blocks_btwn_input_txs_total}
&
Total accumulated block intervals between consecutive occasions on
which the address acts as an input address.
\\

42
&
\feature{blocks_btwn_input_txs_min}
&
Minimum block interval between consecutive input-side appearances.
\\

43
&
\feature{blocks_btwn_input_txs_max}
&
Maximum block interval between consecutive input-side appearances.
\\

44
&
\feature{blocks_btwn_input_txs_mean}
&
Mean block interval between consecutive input-side appearances.
\\

45
&
\feature{blocks_btwn_input_txs_median}
&
Median block interval between consecutive input-side appearances.
\\

46
&
\feature{blocks_btwn_output_txs_total}
&
Total accumulated block intervals between consecutive occasions on
which the address acts as an output address.
\\

47
&
\feature{blocks_btwn_output_txs_min}
&
Minimum block interval between consecutive output-side appearances.
\\

48
&
\feature{blocks_btwn_output_txs_max}
&
Maximum block interval between consecutive output-side appearances.
\\

49
&
\feature{blocks_btwn_output_txs_mean}
&
Mean block interval between consecutive output-side appearances.
\\

50
&
\feature{blocks_btwn_output_txs_median}
&
Median block interval between consecutive output-side appearances.
\\

51
&
\feature{num_addr_transacted_multiple}
&
Number of distinct counterparty addresses with which the focal
address transacted more than once.
\\

52
&
\feature{transacted_w_address_total}
&
Total number of address-to-address interaction records involving
the focal address.
\\

53
&
\feature{transacted_w_address_min}
&
Minimum number of interactions with an individual counterparty
address.
\\

54
&
\feature{transacted_w_address_max}
&
Maximum number of interactions with an individual counterparty
address.
\\

55
&
\feature{transacted_w_address_mean}
&
Mean number of interactions per counterparty address.
\\

56
&
\feature{transacted_w_address_median}
&
Median number of interactions per counterparty address.
\\

\end{longtable}
\endgroup

\section{Definitions of the Graph- and Directed-Hypergraph-Based Address Features}
\label{para:mining_address_graph_hypergraph_features}

In order to describe the graph-based and hypergraph-based features of the data set, we need to introduce the concepts of graph and hypergraph first.

\medskip
\noindent
\textbf{Directed Graph.}
A directed graph is defined as
\[
    G=(V,E),
\]
where
\[
    V=\{v_1,v_2,\ldots,v_N\}
\]
is the set of nodes and
\[
    E\subseteq V\times V
\]
is the set of directed edges. A directed edge
\[
    (v_i,v_j)\in E
\]
points from node \(v_i\) to node \(v_j\).

For a node \(v_i\), the in-degree is the number of directed edges entering
the node:
\[
    d^{\mathrm{in}}(v_i)
    =
    \left|
        \left\{
            v_j\in V:
            (v_j,v_i)\in E
        \right\}
    \right|.
\]
The out-degree is the number of directed edges leaving the node:
\[
    d^{\mathrm{out}}(v_i)
    =
    \left|
        \left\{
            v_j\in V:
            (v_i,v_j)\in E
        \right\}
    \right|.
\]
The total degree is
\[
    d(v_i)
    =
    d^{\mathrm{in}}(v_i)
    +
    d^{\mathrm{out}}(v_i).
\]

\medskip
\noindent
\textbf{Directed Hypergraph.}
A directed hypergraph is defined as
\[
    \mathcal{H}
    =
    \left(
        V,
        \mathcal{E}
    \right),
\]
where \(V\) is the set of nodes and
\[
    \mathcal{E}
    =
    \left\{
        e_1,e_2,\ldots,e_K
    \right\}
\]
is the set of directed hyperedges. Each directed hyperedge is written as
\[
    e_k
    =
    \left(
        T_k,
        H_k
    \right),
\]
where
\[
    T_k\subseteq V
\]
is the tail-node set and
\[
    H_k\subseteq V
\]
is the head-node set. The directed hyperedge represents a higher-order
relationship from the tail set to the head set:
\[
    T_k\longrightarrow H_k.
\]

The in-degree of node \(v_i\) in the directed hypergraph is the number of
directed hyperedges whose head sets contain \(v_i\):
\[
    d_{\mathcal{H}}^{\mathrm{in}}(v_i)
    =
    \left|
        \left\{
            e_k=(T_k,H_k)\in\mathcal{E}:
            v_i\in H_k
        \right\}
    \right|.
\]
The out-degree of node \(v_i\) is the number of directed hyperedges whose
tail sets contain \(v_i\):
\[
    d_{\mathcal{H}}^{\mathrm{out}}(v_i)
    =
    \left|
        \left\{
            e_k=(T_k,H_k)\in\mathcal{E}:
            v_i\in T_k
        \right\}
    \right|.
\]
The total directed-hypergraph degree is
\[
    d_{\mathcal{H}}(v_i)
    =
    d_{\mathcal{H}}^{\mathrm{in}}(v_i)
    +
    d_{\mathcal{H}}^{\mathrm{out}}(v_i).
\]

Thus, the Elliptic++ Bitcoin transaction data set can be represented as a
directed hypergraph. Due to the reasons discussed in the previous section,
we treat each wallet address, also referred to as an actor, as a node and
represent the movement of Bitcoin generated by each transaction as a
directed hyperedge. Since one Bitcoin transaction may transfer funds from
multiple input addresses to multiple output addresses, each transaction is
represented as a many-to-many directed hyperedge whose tail contains the
input addresses and whose head contains the output addresses.

The directed transaction hypergraph is denoted by
\[
    \mathcal{H}
    =
    \left(
        V,\mathcal{E}
    \right).
\]
Each transaction \(t\) is represented as a directed hyperedge
\[
    e_t
    =
    \left(
        T_t,H_t
    \right),
\]
where
\[
    T_t\subseteq V
\]
is the set of distinct input addresses and
\[
    H_t\subseteq V
\]
is the set of distinct output addresses. The direction of the transaction
hyperedge is therefore
\[
    T_t\longrightarrow H_t.
\]
The set of transaction hyperedges incident to address \(a\) is
\[
    \mathcal{E}(a)
    =
    \left\{
        e_t\in\mathcal{E}:
        a\in T_t\cup H_t
    \right\},
\]
and
\[
    h_a
    =
    |\mathcal{E}(a)|
\]
denotes the number of transaction hyperedges incident to address \(a\).

\end{document}